\documentclass[twocolumn]{aastex63}
\usepackage{graphicx}
\usepackage[countmax]{subfloat}
\usepackage{booktabs}
\usepackage{epsfig}
\usepackage{mwe}
\usepackage{placeins}
\usepackage{float}
\usepackage{amsmath}
\usepackage[figuresright]{rotating}
\newcommand{\RNum}[1]{\uppercase\expandafter{\romannumeral #1\relax}}

\begin{document}
\title{Classification of blazar candidates of unknown type in Fermi 4LAC by unanimous voting from multiple Machine Learning Algorithms}

\correspondingauthor{Aditi Agarwal}
\email{aditi.agarwal@rri.res.in}

\author[0000-0003-4682-5166]{A. Agarwal}
\affiliation{Raman Research Institute, C. V. Raman Avenue, Sadashivanagar, Bengaluru - 560 080, INDIA}

\begin{abstract}

The Fermi fourth catalog of active galactic nuclei (AGNs) data release 3 (4LAC-DR3) contains 3407 AGNs, out of which 755 are flat spectrum radio quasars (FSRQs), 1379 are BL Lacertae objects (BL Lacs), 1208 are blazars of unknown (BCUs) type, while 65 are non AGNs. Accurate categorization of many unassociated blazars still remains a challenge due to the lack of sufficient optical spectral information. The aim of this work is to use high-precision, optimized machine learning (ML) algorithms to classify BCUs into BL Lacs and FSRQs. To address this, we selected the 4LAC-DR3 Clean sample (i.e., sources with no analysis flags) containing 1115 BCUs. We employ five different supervised ML algorithms, namely, random forest, logistic regression, XGBoost, CatBoost, and neural network with seven features: Photon index, synchrotron-peak frequency, Pivot Energy, Photon index at Pivot\_Energy, Fractional variability, $\nu F\nu$  at synchrotron-peak frequency, and Variability index. Combining results from all models leads to better accuracy and more robust predictions. These five methods together classified 610 BCUs as BL Lacs and 333 BCUs as FSRQs with a classification metric area under the curve $>$ 0.96. Our results are significantly compatible with recent studies as well. The output from this study provides a larger blazar sample with many new targets that could be used for forthcoming multi-wavelength surveys. This work can be further extended by adding features in X-rays, UV, visible, and radio wavelengths.

\end{abstract}

\keywords{Galaxies: active~-- BL Lacertae objects: general~-- Astronomical Databases: catalogs~-- }

\section{Introduction}
\label{sect:intro}

The launch of the Fermi Gamma-Ray Space Telescope in 2008 started a new era in the identification of $gamma$-ray bright sources. Over the past decade, four Fermi-Large Area Telescope (LAT) source catalogs (FGL) have been published at regular intervals revealing multiple high-energy sources such as AGNs, pulsars, gamma-ray bursts, supernovae, starburst galaxies, etc. The 1FGL catalog released after 11 months has 1451 sources, among which 630 were unassociated ones \citep{2010ApJ...716...30A}; the 2FGL catalog released after two years contains a total of 1873 sources with unassociated ones reduced to 576 \citep{2012ApJS..199...31N}; 3FGL consists of 3033 sources with mostly AGNs and pulsars and about one third were unassociated \citep{2015ApJS..218...23A}. The fourth and latest catalog, 4FGL, is based on the analysis of the 8
yr of LAT data spanning the time range from 2008 to 2016 for the energy range 50 MeV to 1 TeV and contained 5064 sources \citep{2020ApJS..247...33A}. Every FGL catalog was independent of others as each of them was made using new analysis methods, calibrations, diffuse models, and event constructions. Recently, the third release of 4FGL (DR3) contains 6658 sources, of which approximately 2157 are unassociated ones \citep{2022ApJS..260...53A}.

Blazars are a radio-loud type of active galactic nuclei (AGNs) that emit their highly magnetized and relativistic jets within a few degrees along our line of sight \citep{1995PASP..107..803U, galaxies10010006, 2018Galax...6..116R}. Blazars are known for their high variability across the entire electromagnetic spectrum and on multiple timescales ranging from minutes to even decades \citep[e.g.][and references therein]{1995ARA&A..33..163W, 2021A&A...645A.137A}. Depending on their optical emission line properties, blazars are further divided into two categories: Flat Spectrum Radio Quasars (FSRQs) and BL Lacertae objects (BL Lacs). BL Lacs display no or very weak emission line (with equivalent width $<$ 5 $\AA$), whereas on the other hand, FSRQs show wider (stronger) emission lines and a flatter spectrum. Furthermore, BL Lacs are found to have a lower luminosity as compared to FSRQs, thus indicating that different physical mechanisms are occurring in these classes. These flux changes in blazars are further associated with spectral changes \citep{2010ApJ...725.2344B, 2022ApJ...933...42A}. Flux variability can be attributed to various factors such as the Doppler Factor variations, shock formation, variation in the Lorentz factor, injection of new electron population, variation in the magnetic field, and many more. These flux changes can lead to changes in the spectral index of the non-thermal relativistic particle population. The effect of one or more of these parameters can also reflect in their spectral energy distributions (SEDs), like shifts in the synchrotron peak frequency and the corresponding $\nu F_{\nu}$ values. The SED of blazars, i.e., $\nu F_{\nu}$ versus $\nu$ plot, displays a characteristic double hump structure \citep{1998MNRAS.299..433F}. The lower energy hump is located between $10^{13}$ to $10^{17}$ and is typically attributed to the synchrotron emission by relativistic electrons of the jet. The other hump, which is the high energy hump, peaks between 1 MeV to 100 GeV and is generally explained by inverse Compton scattering of UV/Visible/infrared photons by highly energetic particles \citep{2016ARA&A..54..725M}. The soft photons may include the synchrotron photons from the jet itself, and the respective process of inverse comptonization, in this case, is called Synchrotron Self Compton (SSC). Also, the photon field could also be external to the jet environment, e.g., from the accretion disk, broad line region, torus, or Cosmic Microwave Background (CMB) and commonly referred to as External Compton (EC). The second framework popularly used in literature to explain the high energy emission in blazars is based on the hadronic interactions, according to which the higher energy hump is ascribed to hadron-hadron and photon-hadron interactions \citep{2013ApJ...768...54B}. Understanding the origin of the higher energy bump is still a topic of debate. 

Based on the location of the peak for the lower energy SED hump, BL Lacs are further divided into four different classes. BL Lacs with peak frequency below $10^{14}$ Hz are Low-frequency peaked BL Lac objects (LBLs); those with a peak between $10^{14}$ and $10^{15}$ Hz are Intermediate-frequency peaked BL Lac objects (IBL); while the BL Lacs with the peak between $10^{15}$ and $10^{17}$ Hz are High-frequency peaked BL Lac object (HBL); and finally if the peak lies at a frequency more than $10^{17}$ Hz, those BL Lacs are Extreme High-frequency peaked BL Lacs (EHBLs) \citep{2017MNRAS.469..255G}. The Fourth Fermi Gamma-ray LAT (4FGL) source catalog follows the blazar classification scheme as defined by \citet{2010ApJ...716...30A}. According to this, both FSRQs and BL Lacs have been classified into four sub-classes: Low-Synchrotron Peaked (LSP, equivalent to LBL defined above),
Intermediate-Synchrotron Peaked (ISP, similar to IBL class of BL Lacs defined above) and High-Synchrotron Peaked (HSP, equivalent to HBL
and EHBL for BL Lacs).

The fourth catalog of AGNs detected by the Fermi Gamma-ray Space Telescope Large Area Telescope, data release 3 \citep[4LAC-DR3;][]{2022arXiv220912070T} is derived from 4FGL-DR3 \citep{2022ApJS..260...53A}, which is based on 12 years of data and contains about one-third of Blazar Candidates of Uncertain types (BCUs). Classifying these BCUs to BL Lacs and FSRQs will further increase our sample of BL Lacs and FSRQs and thus create a more complete sample of blazars. The rigorous classification of blazars is challenging due to the increased difficulty of obtaining extensive optical spectroscopy observations and also due to a limited understanding of their intrinsic characteristics. Another alternate methodology is based on generating and analyzing multi-wavelength SEDs. But the task of obtaining multi-wavelength observations is time-consuming and thus making it inefficient. For these reasons, it is important to find alternate ways for the classification of blazars to their subclasses. Here, Machine Learning (ML) plays a powerful role in the identification and classification of uncertain-type objects. Many studies have been done using ML algorithms in order to characterize and classify unassociated sources from the Fermi catalogs. Some of them are summarized below: \citet{2012ApJ...753...83A} classified 630 unassociated sources from 1FGL catalog to 221 AGNs and 134 pulsars using RF and LR algorithms, \citet{2012MNRAS.424L..64M} classified 269 high latitude unassociated sources of 2FGL catalog into 216 AGN candidates using RF. Using RF and SVM on 2FGL, \citet{2013MNRAS.428..220H} classified 269 BCUs to BL Lac or FSRQs. \citet{2014ApJ...782...41D} identified AGNs and non-AGNs from 576 unassociated sources of 2FGL using ANN and RF algorithms. After the release of the 3FGL catalog, \citet{2016MNRAS.462.3180C} applied ANN to identify 342 BL Lac objects and 154 FSRQs candidates among the BCUs. Later, \citet{2016Galax...4...14E} used 3FGL along with IR and X-ray data  to search for blazar candidates using the RF method. \citet{2017A&A...602A..86L} identified 3FGL unassociated sources as blazars and further those blazars to 417 BL Lac objects, 149 FSRQs using CNN. \citet{2017MNRAS.470.1291S} applied ANN to identify 271 BL Lac objects and 185 FSRQs from 559 3FGL unassociated sources. \citet{2019A&A...632A..77C} present the largest catalog of HSPs. Using three ML algorithms \citet{2019ApJ...887..134K} predicted 724 BL Lacs and 332 FSRQs from the 4FGL-DR1 catalog. \citet{2020MNRAS.493.1926K} classified 1329 BCUS of 4FGL as 801 BL Lacs, 406 FSRQs using ANN remaining 122 are marked as unclassified. Using a combination of RF and ANN, \citet{2021RAA....21...15Z} studied 1336 unassociated sources and classified them as 583 AGN-like sources, 115 pulsars, 154 other classes, and 484 of uncertain type. \citet{2021ApJ...908..177K} utilize RF to classify Fermi 3FGL unassociated gamma-ray sources as BL Lacs and FSRQs. 84 highly likely blazar candidate sample was then classified as 50 likely BL Lac objects, and 34 were ambiguous.

The nature of numerous $\gamma$-ray sources is still not completely known, and therefore in this work, we revisit the problem of classifying blazar candidates to their sub-classes: BL Lacs and FSRQ, using the 4LAC DR3 catalog by ML-based algorithms, including Random Forest (RF), Logistic Regression (LR), XGBoost, CatBoost, and Neural Network (NN).
Our expected output is to optimize ML algorithms further to classify BCUs from the latest release of the Fermi-LAT source catalog with better accuracy. The paper is organized as follows: in Section 2, we provide a brief description of ML algorithms used for the analysis. Section 3, gives details about data preparation and methodology. In Section 4, we present the output of our models, and finally, in section 5, we discuss our results.

\section{Machine Learning Algorithms}

\subsection{Random Forest}

ML has a plethora of classification algorithms, including RF, LR, Support Vector Machine (SVM), naive Bayes classifier, decision trees, and many more.
In this study, we used the RF method \citep{breiman2001random}, a supervised, ensemble learning, decision-tree-based algorithm for classification and regression. RF is one of the most popular classifier ML algorithms. A group of decision trees is created, and in which each tree is trained on a different subset of the data sample. Unlike the decision tree method, which is built on an entire dataset, using all the features/variables of interest, a random forest is trained by randomly selecting a set of data as well as specific features to build multiple decision trees. After a large number of trees are built using this method, each tree votes or chooses the class. Since RF is a collection of trees, we need to combine their predictions to generate the final prediction of classification to a particular class. For this, we count the times each class is predicted by constituent trees. In the case of binary classification, the class which is predicted the most (majority of the) times is the final prediction of the RF i.e., a majority vote is taken after considering the output from each tree in the forest. Processing the predictions from a large number of relatively uncorrelated models (decision trees) operating as a committee increases the accuracy of the RF even for larger data samples. The low correlation among the models (trees) has been found to increase the success of the model. Apart from high accuracy, the RF scheme also returns classification probabilities along with feature importance rankings.

\subsection{Logistic regression}

Logistic Regression (LR) \citep{hastie2009elements} is a form of a supervised classification algorithm to predict the outcome in the form of true or false (binary dependent variables), unlike linear regression, which is used to predict the continuous value of dependent variables. Similar to linear regression, LR also assumes linearity between the outcome and the variables. The weighted sum of input features (including bias) in LR is sent to a sigmoid function that only predicts output values between 0 and 1.
LR follows the maximum likelihood method to find the best fit line. It maximizes the Maximum Likelihood Estimation (MLE) to determine the parameters and therefore relies on large-sample assumptions. The likelihood in our case is the probability of data points being part of the BL Lac class or FSRQ class of blazars.

\subsection{XGBoost}

XGBoost \citep{2016arXiv160302754C}, an acronym for eXtreme Gradient Boosting, is a tree-based boosting algorithm developed using gradient boosting.
Since its inception in 2016, XGBoost has solved classification and regression problems exceptionally well.
XGBoost is also technically an ensemble algorithm where models are merged together to get a final conclusion but with a more intelligent strategy.
Gradient boosting is a process of sequentially generating decision trees with the objective of minimizing errors.  In ML, boosting originated in computational learning theory with the idea of modifying a weak hypothesis or a weak learner to a stronger and better learner, which finally increases the model's accuracy and performance. Boosting trains all trees in succession and not in isolation such that each new tree is trained such that it minimizes the error made by the previous ones. Trees are then added in a sequential manner to deduce the final prediction by minimizing the loss function. Recently, XGBoost has gained recognition for solving a wide range of astrophysical problems, such as the classification of unassociated 3FGL source samples to their respective classes \citep{2016ApJ...825...69M}, search for quasar candidates \citep{2019MNRAS.485.4539J}, identify some potentially repeating FRBs \citep{2022arXiv221002463L}, and many more.

\subsection{CatBoost}

CatBoost or Categorical Boosting \citep{10.5555/3327757.3327770} is a relatively new ensemble ML algorithm used for not just classification and regression problems but also works well for ranking, forecasting, self-driving car systems and recommendation problems. It is developed by Yandex and is available as an open-source gradient boosting library. CatBoost introduces the concept of ordered boosting to avoid overfitting noisy datasets and prevent target leakages. Ordered boosting is a random multiple permutation approach to train your model on a subset of data with other supporting models also maintained.
Unlike XGBoost, CatBoost creates symmetric trees such that a single tree structure is shared, i.e., leaves from the former trees are trained using the same features. In other words, CatBoost uses oblivious decision trees, one where the same features are used as splitting criteria. Thus it saves a lot of prediction time and is easily scalable for big data sets. All features used in previous splits of the tree are combined with all the categorical features of the data. Since its debut in 2018, CatBoost has been implemented in a number of Big data research \citep{2022MNRAS.515.1807C, 2022MNRAS.509.2289L} and thus illustrates its effectiveness in classification scenarios.

\subsection{Neural Network}
Neural Networks (NN) \citep{Goodfellow-et-al-2016} gained significant popularity in the early 1980s owing to their resemblance in working with biological neurons. Although, the hype was relatively short-lived. NN went into oblivion and gained popularity in the 2010s with increased computational power and the advent of deep learning. It beat existing benchmarks in possibly all of the ML domains, whether it be computer vision, Natural Language Processing(NLP), machine translation, speech recognition, recommendation and personalization, and information retrieval and web search. Currently, almost all of the state-of-the-art (SOTA) algorithms for problems in these domains are based on NN.
The most common form of the NN is a feed-forward network. It consists of multiple layers of neurons. The leftmost layer consists of the input layer {\bf(Fig. \ref{fig:nndiag})}. Each neuron represents a feature input to the model. The rightmost layer is the output layer. Depending on the outcome of the model, the output can have a single or multiple neurons. The middle layers are called the hidden layers. The neurons of two consecutive layers are interconnected with associated weights for each connection. The inputs are connected to the input layer, and the model output is emitted out of the output layer. The NNs are generally trained using the backpropagation algorithm, which back-propagates the loss to all layers and updates the connection weights. Mathematically, these weights of the connections represent the coefficients used to multiply the incoming inputs to a neuron to generate its output. These weights are part of the parameters of a neural network that are tuned using the backpropagation algorithm during the training phase. More complex NN are defined for specific use cases. Recurrent Neural Networks (RNNs) \citep{lipton2015critical} are widely used in time series analysis. Long Short-Term Memory (LSTMs) \citep{hochreiter1997long} and Transformers \citep{vaswani2017attention} are used in training language models for Neural Machine Translation(NMT), Question Answering (Q\&A), text summarization, and various other complex language-based tasks. Convolutional Neural Networks (CNNs) \citep{lecun1995convolutional} find use in Computer Vision (CV) applications such as face detection, object detection, Image captioning, etc. More recently, General Adversarial Networks (GANs) \citep{goodfellow2020generative} have been applied to generate realistic photographs, semantic image-to-text translation, fraud detection, and many other use cases.

\section{Data Preparation and Methodology}
\label{sect:dataprep}

\subsection{Sample Selection}
We selected a sample of sources from the 4LAC-DR3 catalog, which includes 792 known FSRQs, 1458 known BL Lacs, and 1493 BCUs. These include sources from both high and low-latitude samples. For the purpose of analysis in this paper, to improve the accuracy, we further used only the sources from the clean sample (please refer to Section 3.7.3 of \citet{2020ApJS..247...33A}), which reduced our working sample to 670 FSRQs, 1335 known BL Lacs, and 1115 BCUs. Thus, the total data size consists of 3120 sources from the 4LAC-DR3 catalog. Note that, in this analysis, we considered only the Fermi catalog - 4LAC-DR3 without adding any external data from various multi-frequency archives. Observational and instrumental effects can affect the data and, thus, the source distributions, which may further impact our results. But on the other hand, we found that each of our ML algorithms was highly effective in classifying a larger sample of BCUs with very high accuracy.

\begin{figure}
\centering
\includegraphics[width=1.0\linewidth,clip=true]{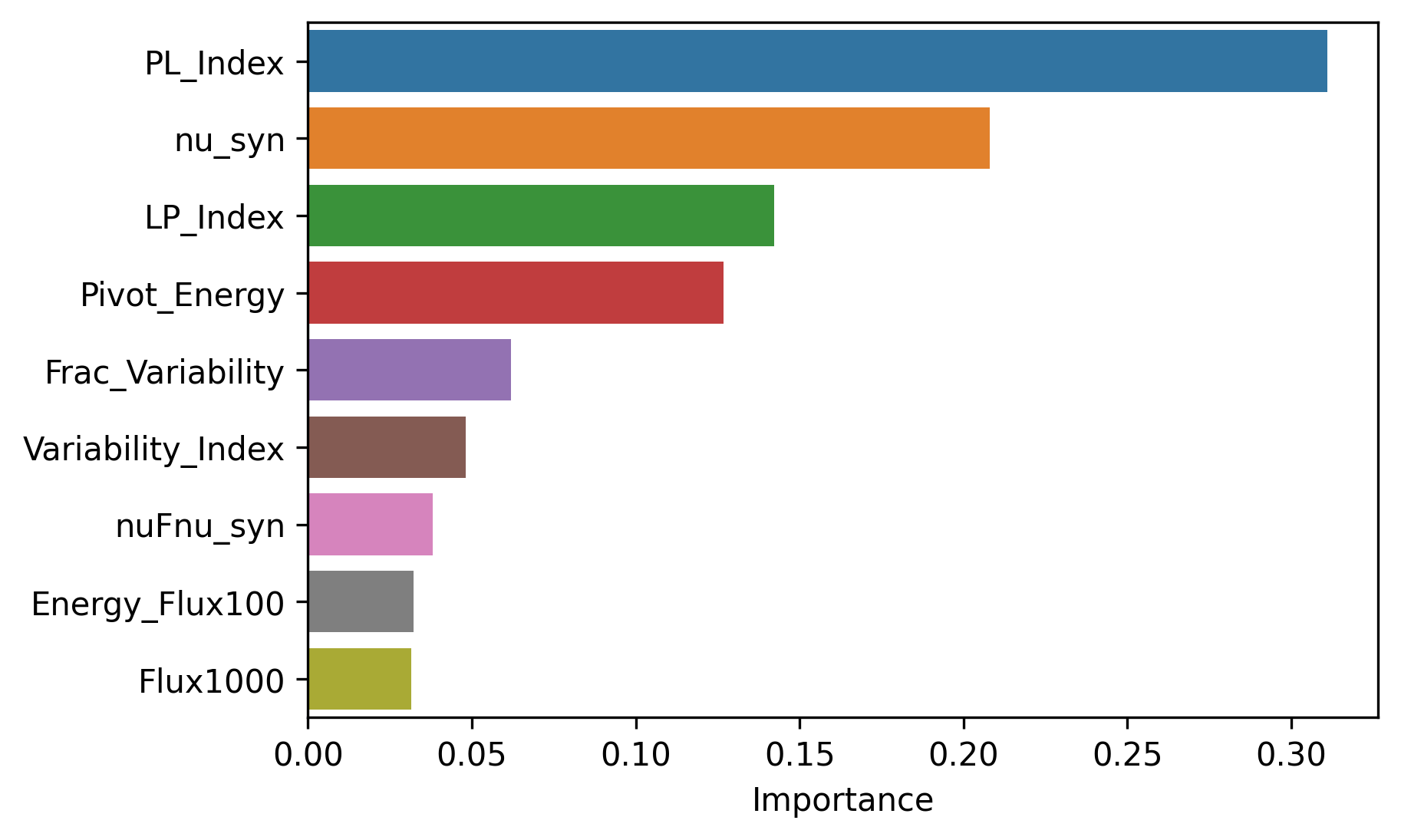}
\caption{\label{fig:allfeaturesimp} Feature importance of all 9 features using RF Classifier}
\end{figure}

\begin{figure*}
\centering
\includegraphics[width=1.0\linewidth,clip=true]{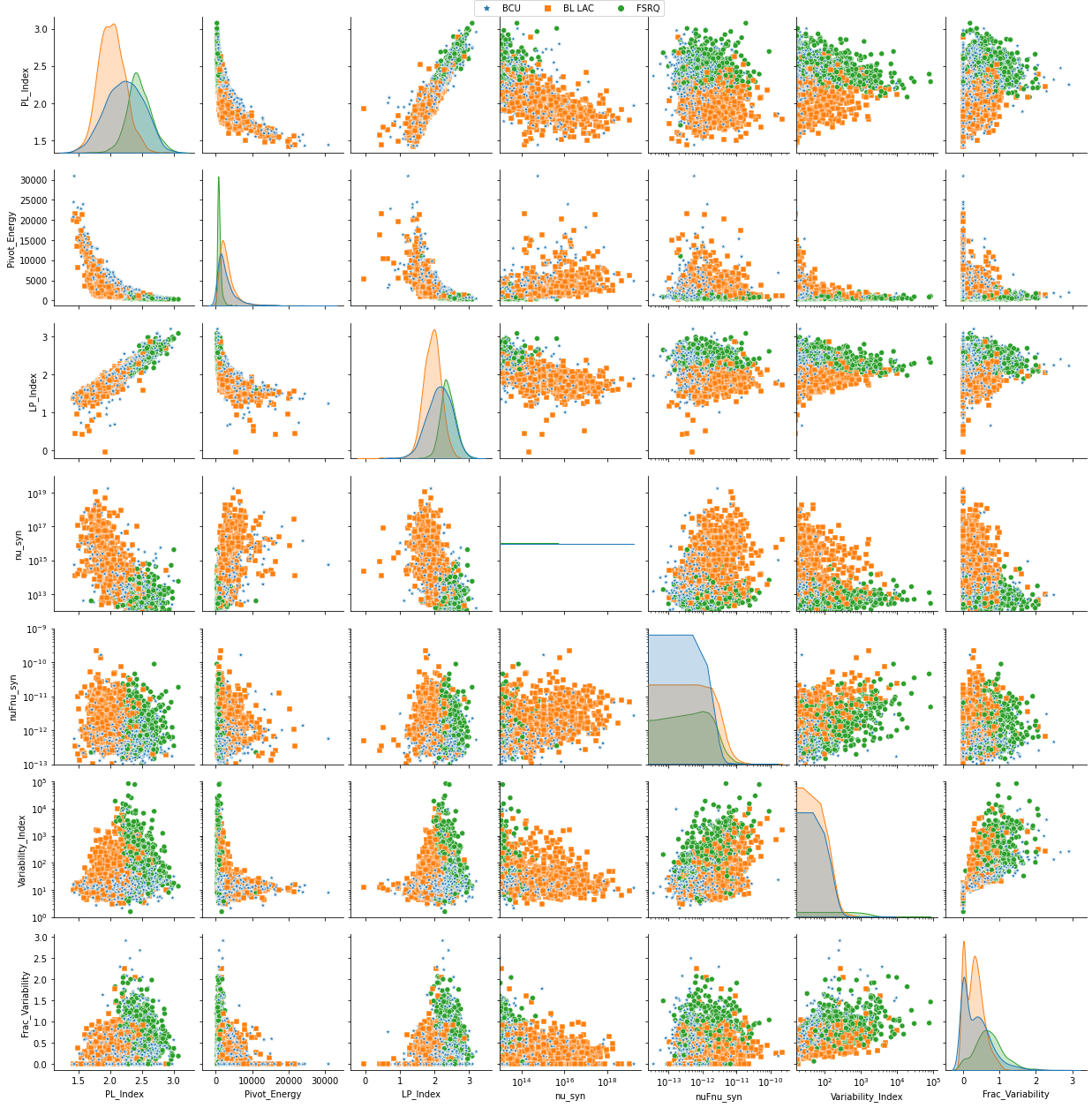}
\caption{\label{fig:pairplot}Pair plots of all features used in training the classifier. We used seven features which are: Photon index when fitting with PowerLaw (PL\_index), Pivot Energy, LP\_Index, nu\_syn, nuFnu\_syn, Frac\_Variability and Variability\_Index. The colors indicate the class of the examples, BCUs (blue), BL Lacs (orange), and FSRQs (green). }
\end{figure*}

We considered all possible features that can be included in our model training. Although, many features had missing values in the original catalog. The feature 'Redshift' was missing for 1540 sources, 'HE\_EPeak' for 442, 'HE\_nuFnuPeak' for 442, and 'Highest\_energy' for 1026 sources. To include these features in the final model, we would have to drop a considerable number of sources from the sample. This would severely impact our model quality and final catalog. Thus, we have removed these features from further consideration. Moreover, it is evident from the 4LAC catalog that most of the known BL Lac candidates are located in the Northern Galactic hemisphere as compared to the Southern Galactic part, which is inevitable because of limited optical spectroscopic data for southern hemisphere candidates. Therefore considering Galactic Latitude as a parameter for ML algorithms is not ideal as BCUs of the Northern hemisphere would then be predicted mostly as BL Lacs only. In addition to these, columns with coordinates, errors, strings, and missing values are also removed. Next, we generated the feature importance metric for all the remaining nine features using the RF classifier. The results are shown in Fig. \ref{fig:allfeaturesimp}. Generally, those features are selected that have more contribution towards the classification output. 
To further remove noisy and irrelevant features, we used the Recursive Feature Elimination with Cross-Validation (RFECV) algorithm for feature selection. In this method, we repeatedly remove the least important feature one step at a time and calculate the impact on model performance by cross-validation. Based on the result of applying this algorithm to our dataset, we eliminated the low-ranked features which were irrelevant. Thus, the features, namely, 'Flux1000' and 'Energy\_Flux100', were removed from further consideration. Notably, this is in agreement with the feature importance plot in Fig. \ref{fig:allfeaturesimp}. Finally, we used seven features which are: Photon index when fitting with PowerLaw (PL\_index), Pivot\_Energy, LP\_Index, nu\_syn, nuFnu\_syn, Frac\_Variability and Variability\_Index. 
To analyze the features further, we plotted the pair-wise scatter plot and class-wise distribution for all the features as shown in Fig. \ref{fig:pairplot}. The plots on the diagonal indicate the class-wise distribution across the feature's range. All other plots are pair-wise scatter plots showing class-wise spread across the features' range. The colors indicate the class of the examples, BCUs (blue), BL Lacs (orange), and FSRQs (green). A detailed examination of the diagonal distribution plots reveals that there is a clear distinction between the sources belonging to class 'BL Lac' vs class 'FSRQ' for features like the PL\_index), Pivot\_Energy, LP\_Index, and Frac\_Variability. The remaining features have a considerable overlap across classes which makes them less important for the classification task. This is an important observation as there is a strong correlation between the separation of class means and its corresponding feature importance in Fig. \ref{fig:allfeaturesimp}. Also, the sources of class BCU are mainly concentrated right in the middle of the two classes, indicating it is a good mix of sources from both classes.

A brief description of these seven features used in this study is given below. For a more detailed understanding, we refer the reader to \citet{2022ApJS..263...24A} and references therein. The spectral analysis in 4LAC-DR3 \citep{2022ApJS..263...24A} has been done following a similar procedure as described in 4FGL-DR1 \citep{2020ApJS..247...33A} except that now a different parameterization is being used for pulsars, more number of sources are fit, the threshold for considering spectral curvature as significant has been lowered, a new column with peak energy in $\nu F_{\mathrm{\nu}}$ has been reported and a spectral bin has been added to SEDs. The spectral representation of sources still follows a power law, power law with subexponential cutoff, and log-normal. The normalization (flux density $K$) for these spectral representations is defined at a reference energy $E_0$ such that the error on $K$ is minimum. $E_0$ appears as \texttt{Pivot\_Energy} in the catalog. A log-normal function is given as:
\begin{equation}
\frac{{\rm d}N}{{\rm d}E} = K \left (\frac{E}{E_0}\right )^{-\alpha -
\beta\log(E/E_0)}.
\label{eq:logparabola}
\end{equation}
The parameters $K$, $\alpha$ (spectral slope at $E_0$) and the curvature $\beta$ appear as \texttt{LP\_Flux\_Density}, \texttt{LP\_Index} and \texttt{LP\_beta} in the catalog, respectively. A more stable parameterization (\texttt{PLSuperExpCutoff4} in the Fermi Tools) used in 4FGL-DR3 is given as:
\begin{eqnarray}
\frac{{\rm d}N}{{\rm d}E} & = & K \left (\frac{E}{E_0}\right )^{\frac{d}{b}-\Gamma_S} \exp \left [\frac{d}{b^2} \left (1 - \left (\frac{E}{E_0}\right )^b \right ) \right ]
\label{eq:expcutoff} \\
\frac{{\rm d}N}{{\rm d}E} & = & K \left (\frac{E}{E_0}\right )^{-\Gamma_S-\frac{d}{2}\ln\frac{E}{E_0}-\frac{db}{6}\ln^2\frac{E}{E_0}-\frac{db^2}{24}\ln^3\frac{E}{E_0}} {\rm for} \left | b \ln\frac{E}{E_0} \right | < 10^{-2},
\label{eq:expcutoff2}
\end{eqnarray}
Here the normalization $K$ is directly the flux density at the reference energy $E_0$ and the shape parameters are the spectral slope $\Gamma_S$ and the spectral curvature $d$ which are defined in \citet{2022ApJS..260...53A}. With this parameterization, the correlation between parameters is reduced considerably. Moreover, as shown by \citet{2022ApJS..260...53A}, $\Gamma_S$ is much better defined (error ratio below 1) than the low-energy index $\Gamma = \Gamma_S - d/b$ used in 4FGL-DR1. The parameters $K$, $\Gamma_S$, $d$ and $b$ appear as \texttt{PLEC\_Flux\_Density}, \texttt{PLEC\_IndexS}, \texttt{PLEC\_ExpfactorS} and \texttt{PLEC\_Exp\_Index} in the FITS table, respectively. Finally, a simple power-law form for all sources not significantly curved is used which is defined as:
\begin{equation}
\frac{{\rm d}N}{{\rm d}E} = K \left (\frac{E}{E_0}\right )^{-\Gamma} 
\end{equation}
For those parameters $K$ and $\Gamma$ appear as \texttt{PL\_Flux\_Density} and \texttt{PL\_Index} in the catalog table. The spectral parameters are reported by associating them with the spectral model they come from, i.e., in the form Shape\_param where param is the parameter name and Shape is PL for PowerLaw, PLEC  for PLSuperExpCutoff, or LP for LogParabola. In addition to that, all sources were fit with all three spectral shapes thus all enteries were filled. Another important property of blazars is that they can also be classified using SED based method which uses the value of peak frequency $\nu_\mathrm{s,peak}$ of the synchrotron component of SEDs. In 4LAC-DR1 \citep{2020ApJ...892..105A}, broadband SEDs for all 4LAC AGNs were generated using the SED-Builder interactive tool available at the Italian Space Agency (ASI) Space Science Data Center (SSDC)\footnote{http://tools.ssdc.asi.it/SED/}. The estimation of $\nu_\mathrm{s,peak}$ and the corresponding $\nu F{_\nu}$ at synchrotron-peak frequency ($erg cm^{-2} s^{-1}$) have been done using two different approaches. One is similar to that followed in previous catalogs (1LAC, 2LAC) i.e., empirical parametrization based on the broadband indices $\alpha_{ro}$  (radio-optical) and $\alpha_{ox}$  (optical-X-rays) \citep{2010ApJ...715..429A,2011ApJ...743..171A}. Whereas the other method more favored in 4LAC-DR1 is the same as that used in 3LAC \citep{2015ApJ...810...14A} and relies on fitting a third-degree polynomial on the low-energy hump of the SEDs. As the measured X-ray flux is not required, more number of sources were assigned $\nu_\mathrm{s,peak}$ value using the second methodology. This fit also gave the $\nu F_{\mathrm{\nu}}$ value at the peak position. Since we do not have redshift information for a large number of 4LAC sources, the frequency in the observer frame was used.  More details on this method can be found in \citet{2015ApJ...810...14A} and \citet{2020ApJ...892..105A}.
Blazars or radio loud AGNs have been found to be variable on diverse timescales and is wavelength dependent. Variability aids in understanding the dominant emission mechanism and dynamics of blazars \citep{2019MNRAS.488.4093A}. In 4FGL, the variability index, $TS_{\rm var}$, is defined as twice the sum of the log(Likelihood)
difference between the flux fitted in each time interval and the average flux over
the full catalog interval \citep{2020ApJS..247...33A} i.e.,
\begin{eqnarray}
TS_{\rm var} & = & 2\sum_i \log \left[ \frac{\mathcal{L}_i(F_i)}{\mathcal{L}_i(F_{\rm glob})} \right] - \max \left( \chi^2(F_{\rm glob}) - \chi^2(F_{\rm av}), 0 \right)
\label{eq::VarIndex} \\
\chi^2(F) & = & \sum_i \frac{(F_i - F)^2}{\sigma_i^2}
\end{eqnarray}
where $F_i$ are the individual flux values, the average flux from the light curve is $F_{\rm av}$, $F_{\rm glob}$ is the flux in the total analysis, $\mathcal{L}_i(F)$ the likelihood in interval $i$ with an assumption that flux $F$ and $\sigma_i$ the errors on $F_i$ (upper error if $F_i \le F$, lower error if $F_i > F$). For 4FGL-DR3, light curves for 12 years of science data, with events up to 1 TeV, were recomputed over 1 year bins following the procedure same as in 4FGL-DR1 \citep{2020ApJS..247...33A}. The threshold for $TS_{\rm var}$ in 4FGL-DR3 is considered to be 24.725 (corresponding to 99\% confidence for 12 intervals) which resulted in increase in number of sources with significant variability from 1443 to 1695.
Whereas, fractional variability for each source is derived from excess variance on top of the statistical and systematic fluctuations and can be calculated as:
\begin{eqnarray}
Var & = & \frac{1}{N_{\rm int}-1} \; \sum_i (F_i - F_{\rm av})^2
\label{eq:samplevar} \\
\delta F & = & \sqrt{\max \left( Var
              - \frac{\sum_i \sigma_i^2}{N_{\rm int}}, 0 \right)}
\label{eq:relvar} \\
\frac{\sigma_F}{F} & = & \max \left( \frac{1}{\sqrt{2(N_{\rm int}-1)}}
              \frac{V_i}{F_{\rm av} \; \delta F }, 10 \right)
\label{eq:uncrelvar}
\end{eqnarray}
Here, fractional variability is defined as $\delta F/F_{\rm av}$ with $\sqrt{N_{\rm int}-1}$ degrees of freedom. The distribution of fractional variability in DR3 was found to be similar to DR1 peaking between 50\% -- 90\%. All parameters used in 4LAC-DR3 are listed in Table A1 of \citet{2022ApJS..263...24A}.

\begin{figure}
\centering
\includegraphics[width=1.0\linewidth,clip=true]{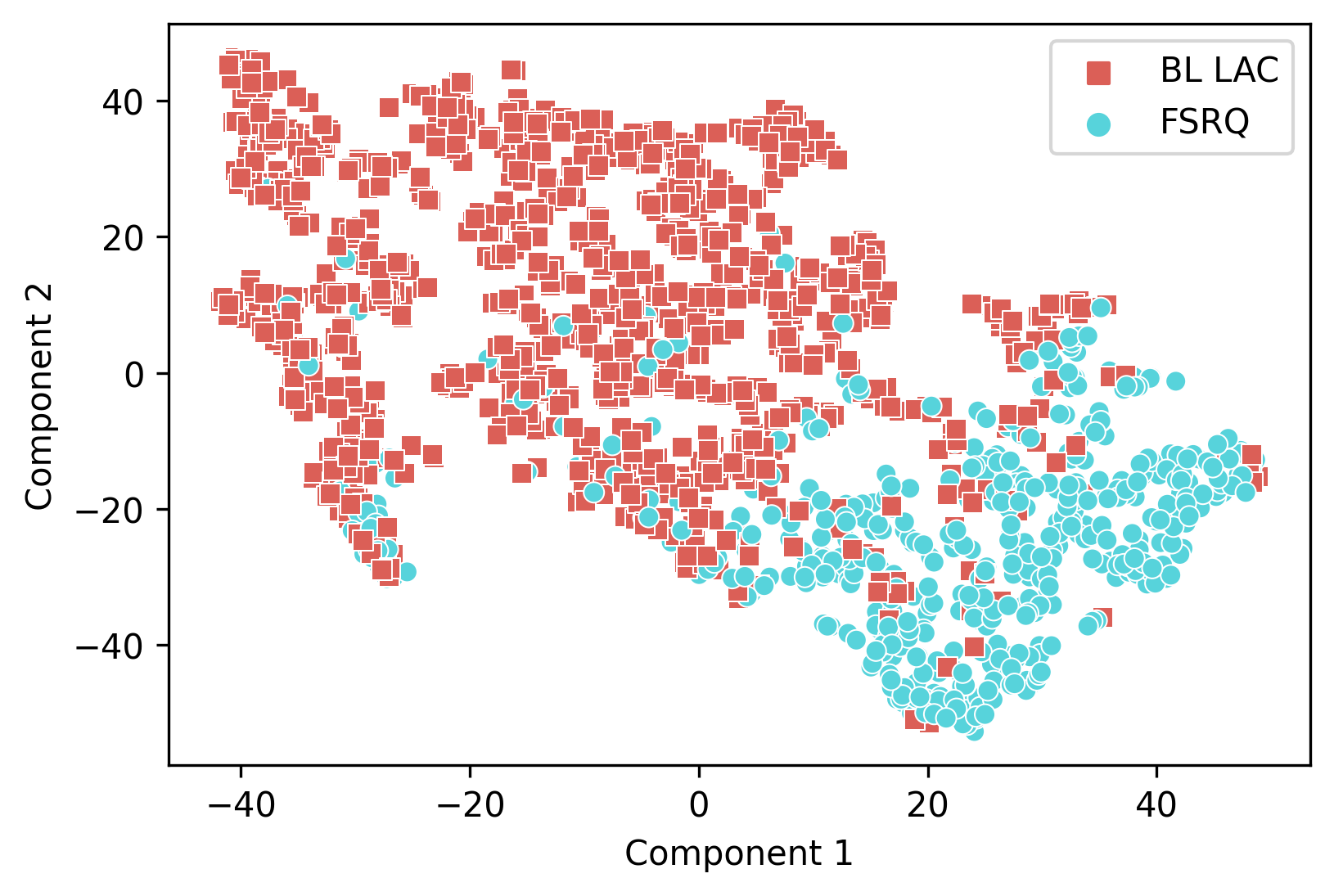}
\caption{\label{fig:tsne}t-SNE plot of the dataset using the top 2 components. It indicates the dataset is reasonably separable wrt. the features used. }
\end{figure}

To further our confidence in the selected features for the task of classification, we ran the T-distributed Stochastic Neighbor Embedding (t-SNE) algorithm over our dataset. t-SNE is a statistical tool to visualize high-dimensional data in a compressed two- or three-dimensional map. The 2 components (called Component-1 and Component-2 in Fig. \ref{fig:tsne}) represent a 2-dimensional subspace onto which the actual high-dimensional data is projected for visualization or representation purposes. The t-SNE plot is shown in Fig. \ref{fig:tsne}. This indicates that the features used are powerful enough to classify the data into two distinct and separable classes.

The first crucial and most important step for ML algorithms is the division of the known sample of BL Lacs and FSRQs into {bf training and testing subsets.} 
Training subset is used by the algorithms to learn each feature/pattern and respective classification class (in our case BL Lac or FSRQ) of all sources with an aim to minimize the loss function. The other subset of data, known as test data, is selected to test the model once training is complete (using training samples) and classification results have been generated. The testing set is carefully sampled in a manner that spans all possible classes which a model would face. It thus provides an unbiased idea of how accurately the model will perform on the unseen dataset. We also used a 5-fold cross-validation (CV) technique to tune the model hyperparameters. In this technique, we divide the complete dataset into five folds or partitions. In the first iteration, the first 4 folds, or 80\% data, are used for training, and the last fold, or 20\% data, is used for evaluation/validation. In subsequent iterations, each fold gets a chance to become the validation set, with the rest of the folds used for training purposes. Hence, all data points get a chance to be part of the validation set exactly once. This enables the comparison of the model performance on various hyperparameter values across multiple train/validation splits.

To assess the accuracy of ML Algorithms, a Receiver Operating Characteristic curve (ROC) is generated for all five algorithms. The ROC graph shows a trade-off between the true positive rate and the false positive rate at different thresholds. The True Positive Rate (TPR) signifies the proportion of positive class samples that are correctly predicted by the model, while the False Positive Rate (FPR) is the measure of the proportion of class samples incorrectly predicted to be positive. 

Finally, we combine the outputs from all five methods; that is, we classify an unassociated source as BL Lac or FSRQ only when it is classified as the same source by all five ML algorithms.

\subsection{Data Preprocessing and Models}

Firstly, we split our final dataset into two parts - The Training set and the Test set. Following \citet{2022arXiv220709307G}, we performed a 5-fold cross-validation (CV) with 20 repetitions on the dataset. In each iteration, we took 80\% of data for the training set, and the remaining 20\% was kept aside as a test set. A complete run of CV would generate 5 test sets without any repetition of data. We repeated this process 20 times with different random seeds to generate 100 training and test sets on our dataset. We used the \texttt{RepeatedStratifiedKFold} package defined in $sklearn$ $v1.0.2$ library \citep{scikit-learn}. Therefore, we had 1068 BL Lacs and 536 FSRQs for the training dataset and 267 BL Lacs and 134 FSRQs for the test dataset. Next, we applied feature normalization to the training set to ensure the ranges of values across features were similar. Some of the algorithms, such as LR and NN, are susceptible to large variations in feature ranges which affect the training process. In contrast, tree-based methods are not affected by the presence of such features. This difference is due to the way their mathematical formulations are defined and minimize the loss functions. We applied the standard normalization technique to replace the features ($X$) with their normalized values ($X_{new}$).

\begin{equation*}
    X_{new} = \frac{(X - \mu_X)}{\sigma_X}
\end{equation*} 

 Here, $\mu_X$ and $\sigma_X$ are the mean and standard deviation of the feature $X$ in the training set. It is evident from the class-wise split of the training set that BL Lacs outnumber FSRQs by 2:1. This would hamper the learning and prediction of FSRQ class due to lower representation.  There are a number of techniques, such as under-sampling, over-sampling, cost-sensitive learning, and Synthetic Minority Oversampling Technique (SMOTE), that can be applied to overcome this imbalance in the dataset. Under-sampling is generally avoided as it reduces the amount of data used while training, thereby considerably affecting the model performance. Over-sampling and cost-sensitive learning have a similar impact in terms of improving model performance by boosting the under-represented samples. SMOTE \citep{Chawla_2002} is also a well-known technique for data augmentation. Although, as observed by \citet{LOPEZ20126585}, cost-sensitive learning outperforms SMOTE in many cases. Also, we saw a detrimental effect on model performance while applying SMOTE. Similar results are also reported by \citet{2021RAA....21...15Z}. Hence, for this study, we will apply cost-sensitive learning as the method to tackle the imbalance in the training dataset.

Finally, we applied various ML algorithms such as LR, NN, RF, XGBoost, and CatBoost on the dataset.
We used the RandomForestClassifier, which is part of the $sklearn$ $v1.0.2$ library \citep{scikit-learn} in python, for training the RF model. We used a 5-fold CV technique to tune the model hyperparameters. We found that generating 100 decision trees splitting till a leaf node has no more than two sources gave us the best performance. Also, we used 'entropy' as the split criterion. For a node $m$ in the decision tree having $n_m$ examples from K classes, we define the proportion $p_{k}$ of examples of class $k \in {0, 1, ..., K-1}$ as

\begin{equation*}
p_{k} = \frac{1}{n_m} \sum I(y = k)
\end{equation*}

 where $I(y = k)$ denotes the indicator function, which is equal to 1 when $(y = k)$ and 0 otherwise.
This proportion can be viewed as the probability of an example belonging to a class $k$ being in node $m$. Hence, we can calculate the entropy, $H(m)$, of this node using the equation:

\begin{equation*}
H(m) = - \sum_k p_{k} \log(p_{k})
\end{equation*}

The algorithm chooses the best split, which gives the most reduction in entropy at each successive level. We also used cost-sensitive learning to improve the classification for the minority class. We enabled this by applying the class\_weight = 'balanced' option while training the model. 

\begin{table*}
\caption{List of parameters used for training various ML models. All other params for the models are set to default value}            
\label{tab:modelopts} 
\centering
\begin{tabular}{cl}
\hline\hline
Model & Training Parameters \\
\hline
Random Forest & \verb|RandomForestClassifier(n_estimators=100, criterion="entropy", min_samples_split=2, |\\ 
& \verb|max_features = "sqrt", bootstrap = True, class_weight= "balanced", max_samples=None)| \\
Logistic Regression & \verb|LogisticRegression(penalty='l2', class_weight= "balanced"| \\
XGBoost Classifier & \verb|XGBClassifier(n_estimator=100, max_depth=3, learning_rate=0.1, objective='binary:logistic')| \\
CatBoost Classifier & \verb|CatBoostClassifier(learning_rate=0.1, depth=6, n_estimators=100)|\\
Neural Network & \verb|model = Sequential(name="Blazar Classification Model") |\\
& \verb|model.add(Input(shape=(7,), name='Input-Layer')) |\\
& \verb|model.add(Dense(64, activation='relu', name='Hidden-Layer-1'))|\\
& \verb|model.add(Dropout(0.5))|\\
& \verb|model.add(Dense(32, activation='relu', name='Hidden-Layer-2'))|\\
& \verb|model.add(Dense(1, activation='sigmoid', name='Output-Layer'))| \\
\hline
\end{tabular}
\end{table*}

We used the LR, part of the $sklearn$ $v1.0.2$ library in python, to train the LR model. Although its name is a misnomer, LR is actually a classification algorithm belonging to the family of linear models. In this model, we use the logistic or the sigmoid function to represent the class-conditional probabilities.

\begin{equation*}
sig(z) = \frac{1}{1+exp(-z)}
\end{equation*}

 The LR model training consists of learning the weight vector $ w = [w_0, w_1, ..., w_n]$ and a bias term $b$. The final classification is represented by 
 
 \begin{equation*}
y = \begin{cases}
    1 & \text{if }  sig(w^TX + b) \geq 0.5 \\
    0 & \text{if }  sig(w^TX + b) < 0.5 \\
    \end{cases}
\end{equation*}

We used the L2 regularization to avoid overfitting to improve the model performance. We also used cost-sensitive learning to improve the classification for the minority class. We enabled this by applying the class\_weight = 'balanced' option while training the model. 

We also trained an XGBoost classifier on the dataset using the $xgboost$ python package. This package provides a sklearn-compatible Python Application Programming Interface (API) for training XGBoost models. We used the XGBClassifier model for the same. We used a 5-fold CV technique to optimize the hyperparameters of the model. We got the best performance by training 100 estimators capping each tree's max depth at 3. We set the learning rate to 0.1 while training. The learning rate controls the shrinkage applied at each successive tree generated. A low value helps in better convergence in general, although taking longer to reach the optimal state. As before, we applied cost-sensitive learning to improve model performance. We manually generated the 'balanced' sample weights and assigned them to the training set. We picked the objective to be 'binary:logistic' as we are working on a two-class problem. 

Next, we trained a CatBoost classifier on the dataset using the $catboost$ python package. CatBoost is also a gradient-boosting algorithm similar to XGBoost. CatBoost works equally well with numerical and categorical features. The major difference is that CatBoost only creates symmetric trees, i.e., the same split is applied at all the nodes on the same level. This reduces the prediction time and also works as a regularizer to avoid overfitting. As before, we applied cost-sensitive learning to improve model performance. We manually generated the 'balanced' sample weights and assigned them to the training set.

\begin{figure}
\centering
\includegraphics[width=1.0\linewidth,clip=true]{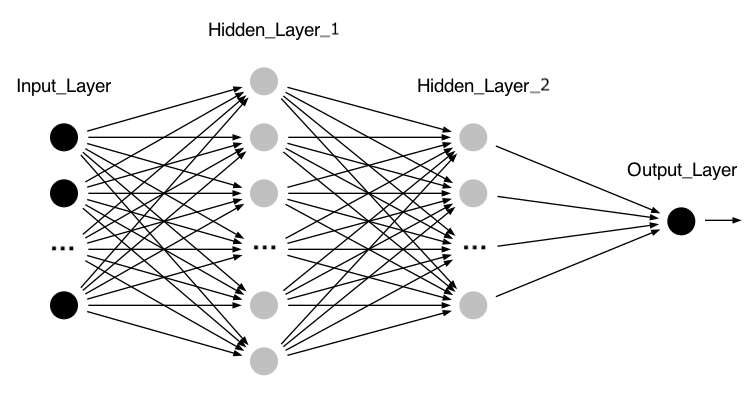}
\caption{\label{fig:nndiag}The network architecture used for training the feed-forward NN. It has 7 neurons in the input layer, one for each input. The first hidden layer has 64 neurons, while the second one has 32 neurons. There is a single neuron in the output layer. }
\end{figure}

Finally, we trained a NN classifier. We used a feed-forward network owing to the tabular nature of our dataset. We started with a basic single hidden layer network and successively added more neurons and additional layers. We used a 5-fold CV technique to tune the model hyperparameters. The best-performing network was a 2-hidden layer network, as shown in Fig. \ref{fig:nndiag}. We also used dropouts for regularization. Dropouts work by randomly dropping neurons from layers during training to add randomization to the network. We added dropouts between 'hidden layer 1' and 'hidden layer 2'. The dropout rate was set to 0.5, i.e., every neuron in the second hidden layer had a 50\% of dropping out. The $relu$ activation function is used in hidden layers, and the $sigmoid$ activation function is for the output layer. We used the $adam$ optimizer as it is very effective in adapting the learning rate. We used "binary cross-entropy" as the loss function since we had a 2-class classification task at hand.

The above set of algorithms forms a good coverage of advanced machine-learning techniques for the classification task. LR is a straightforward binary classifier. RF, CatBoost, and XGBoost are tree-based ensemble methods, whereas NN used in this study is a feed-forward neural network. RF employs bagging techniques, whereas CatBoost and XGBoost are based on gradient boosting. It is to be noted that each of the techniques has its advantages and disadvantages. Therefore, we intend to combine the classification results from all the methods to increase the reliability of predictions and to ignore the predictions with lower confidence.

We have listed all the parameters used for training the various ML models for this work in Table \ref{tab:modelopts}. This would enable researchers to reproduce the results presented in this paper, learn these models, and apply them across other classification/learning tasks.

\section{Results}
\label{sect:res}
Following \citet{2022arXiv220709307G}, we did a 5-fold CV with 20 repetitions on the dataset. In each iteration, the test set consists of 401 sources having 267 BL Lacs and 134 FSRQs. A complete run of CV would generate 5 test sets without any repetition of data. We repeated this process 20 times with different random seeds to generate 100 training and test sets on our dataset. We report the mean performance metrics along with their statistical uncertainty in Table \ref{tab:modelperf}.

The results of the various trained models on the test set are listed in Table \ref{tab:modelperf}. In terms of accuracy, RF, CatBoost, and XGBoost Classifier all gave ~91\% accuracy. NN and LR both had an accuracy of around 89-90\%, which is only marginally lower than the others. The precision for class BL Lac was highest for XGBoost at 94.7\%. The recall for class BL Lac was very high, between 89-96\% for all classifiers. Similarly, the F1-score was very high, between 0.918-0.935 for all classifiers indicating high scores for both precision and recall. Considering the performance for the class FSRQ, the precision of RF, CatBoost, and XGBoost Classifier was ~88\%. LR gave a high recall of 90.1\% for the class FSRQ. Overall, the F1-scores reported were also high in the range of 85-87\%, indicating strong performance for the class FSRQ. We observe very high accuracy, precision, and recall scores, with small uncertainties implying the robustness of the results of each individual classifier.

\begin{figure*}[t!]
\gridline{\fig{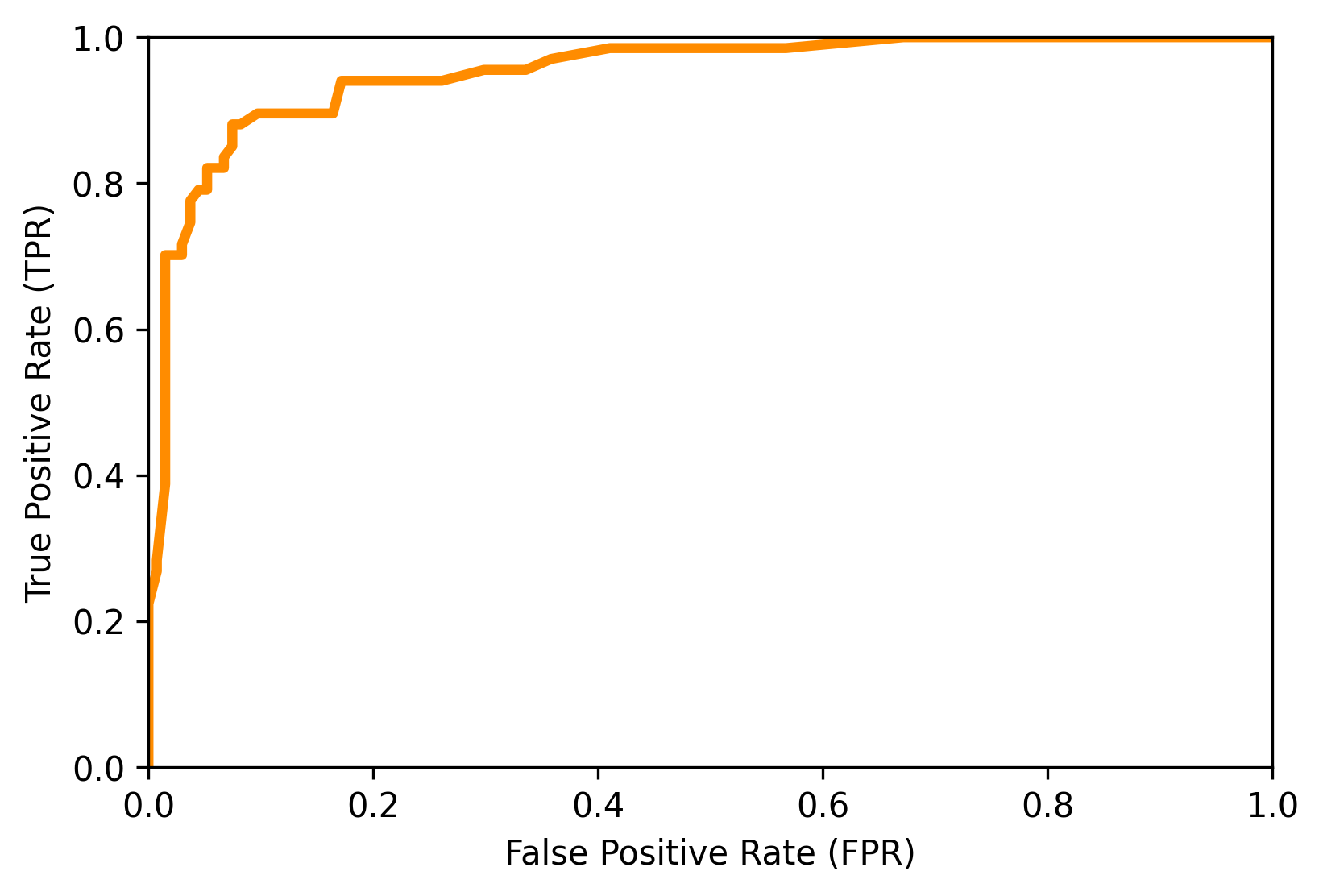}{0.47\textwidth}{\hspace{0.75cm}Random Forest}
          \fig{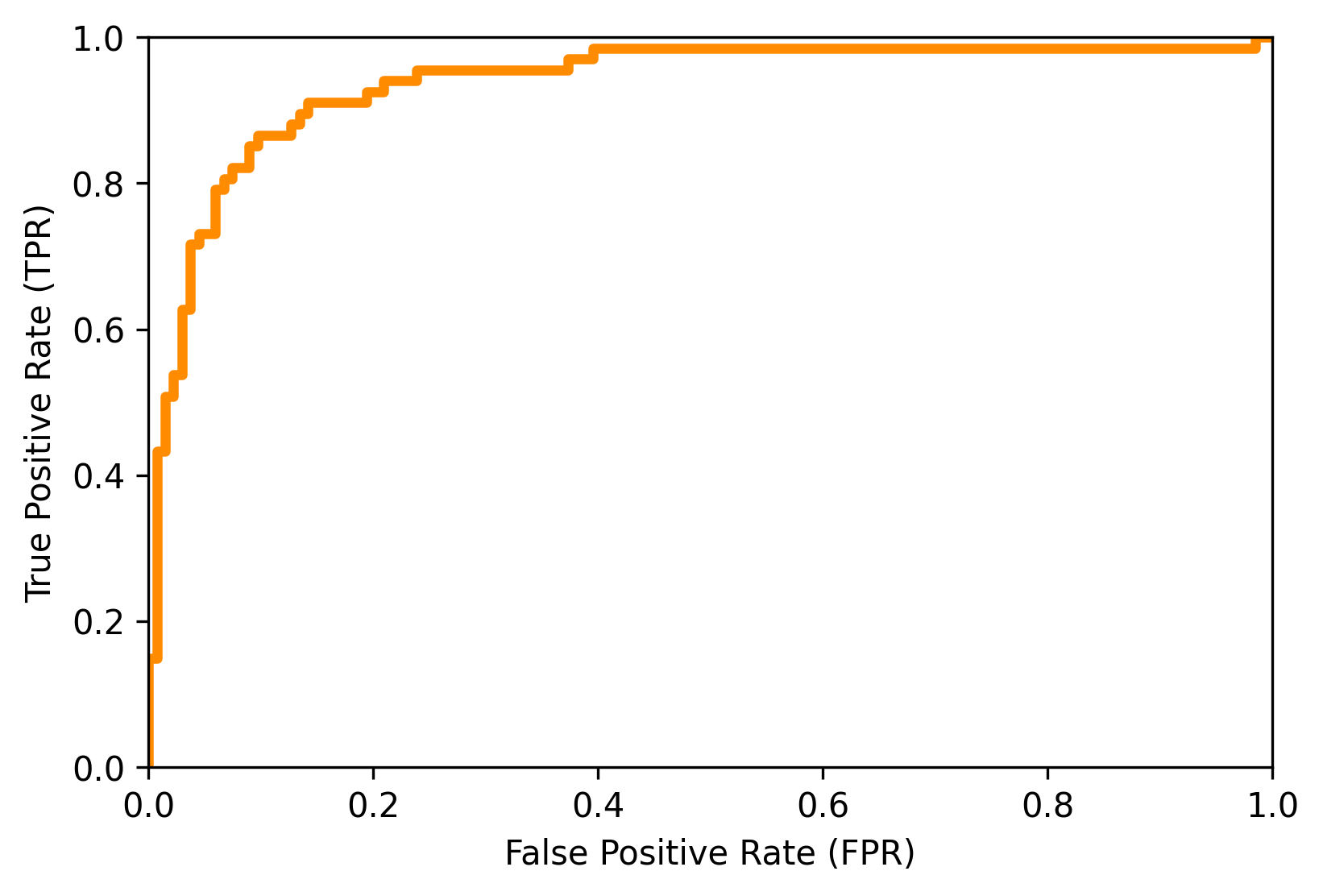}{0.47\textwidth}{\hspace{0.75cm}Logistic Regression}}
\gridline{\fig{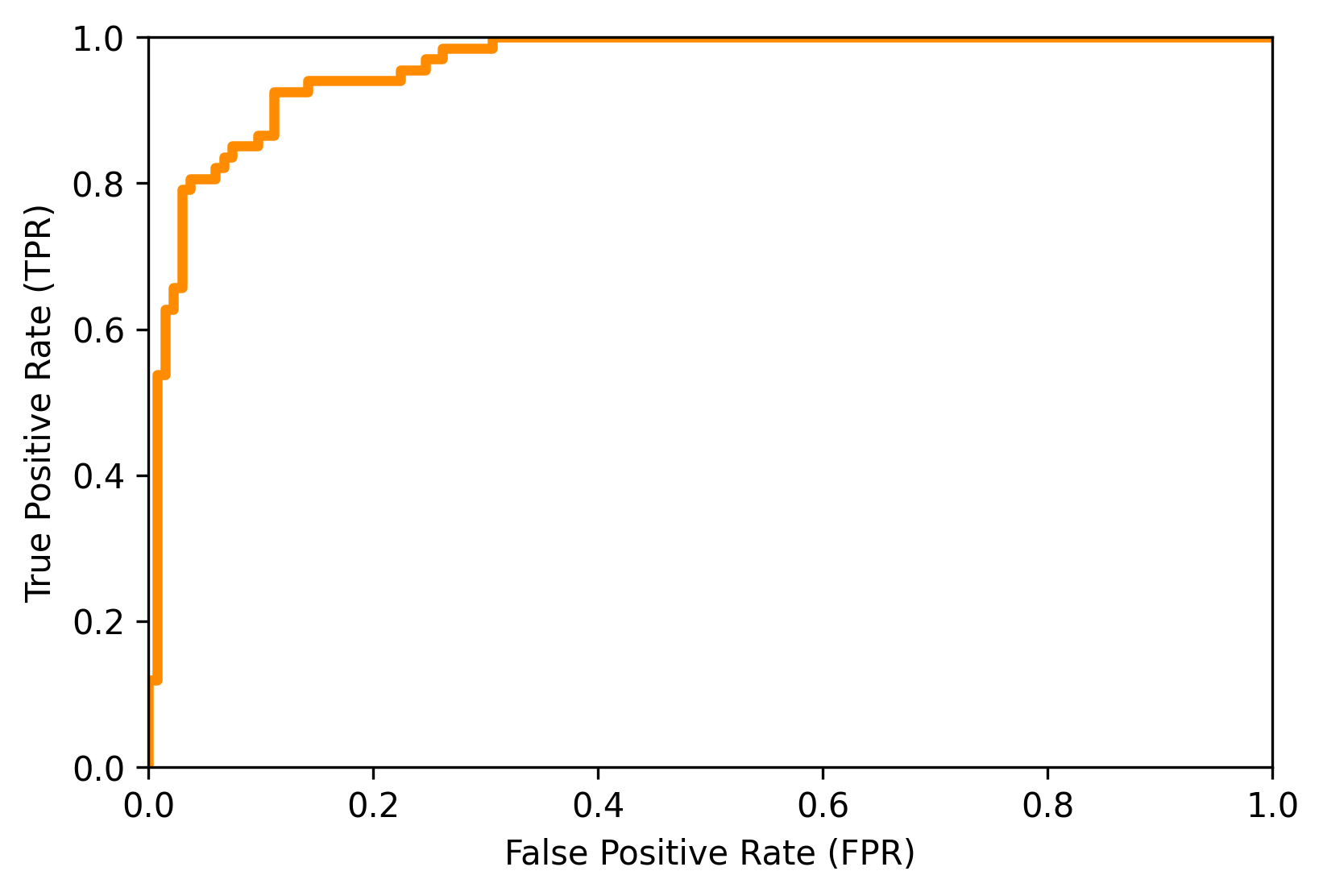}{0.47\textwidth}{\hspace{0.75cm}XGBoost Classifier}
          \fig{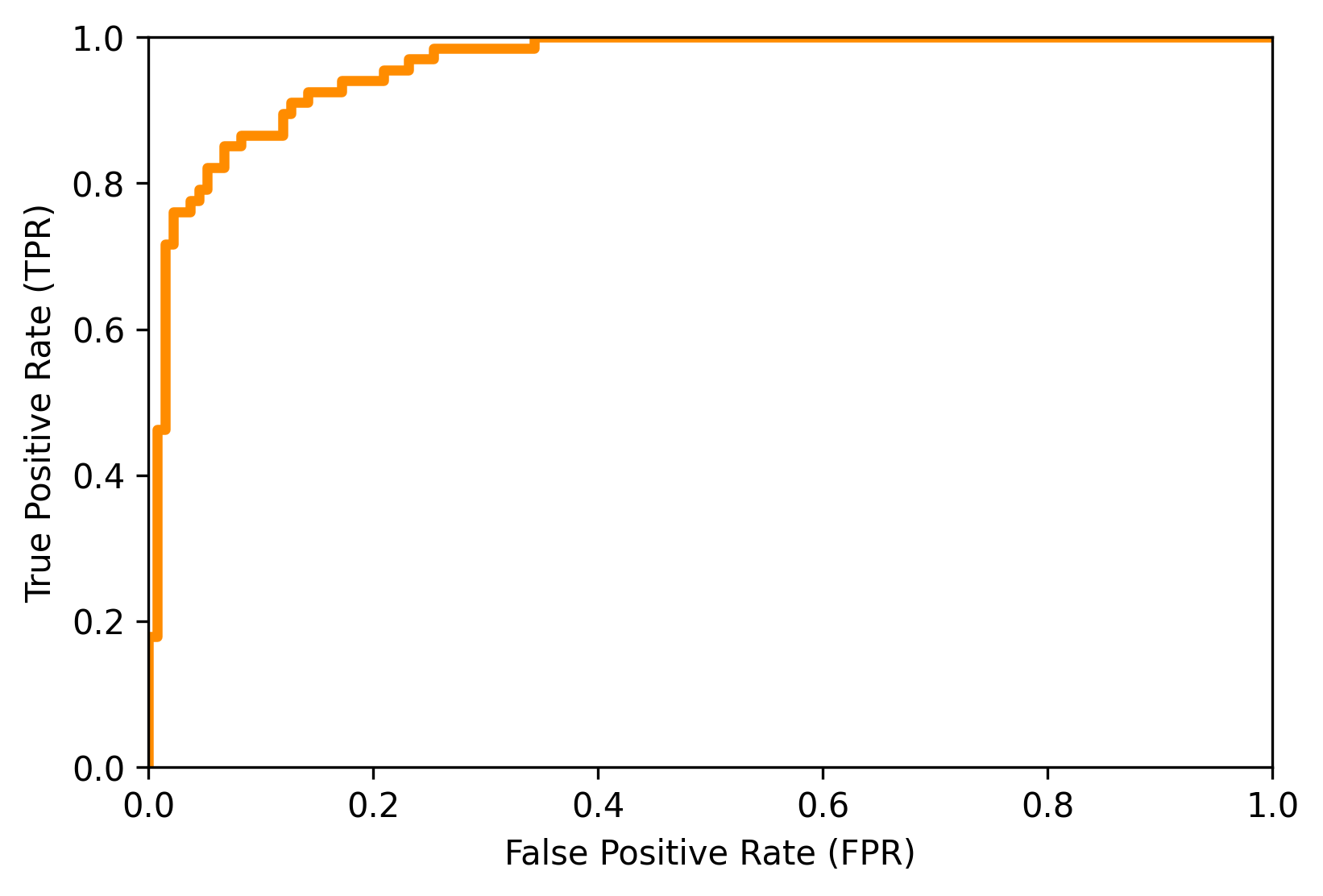}{0.47\textwidth}{\hspace{0.75cm}CatBoost Classifier}}
\gridline{\fig{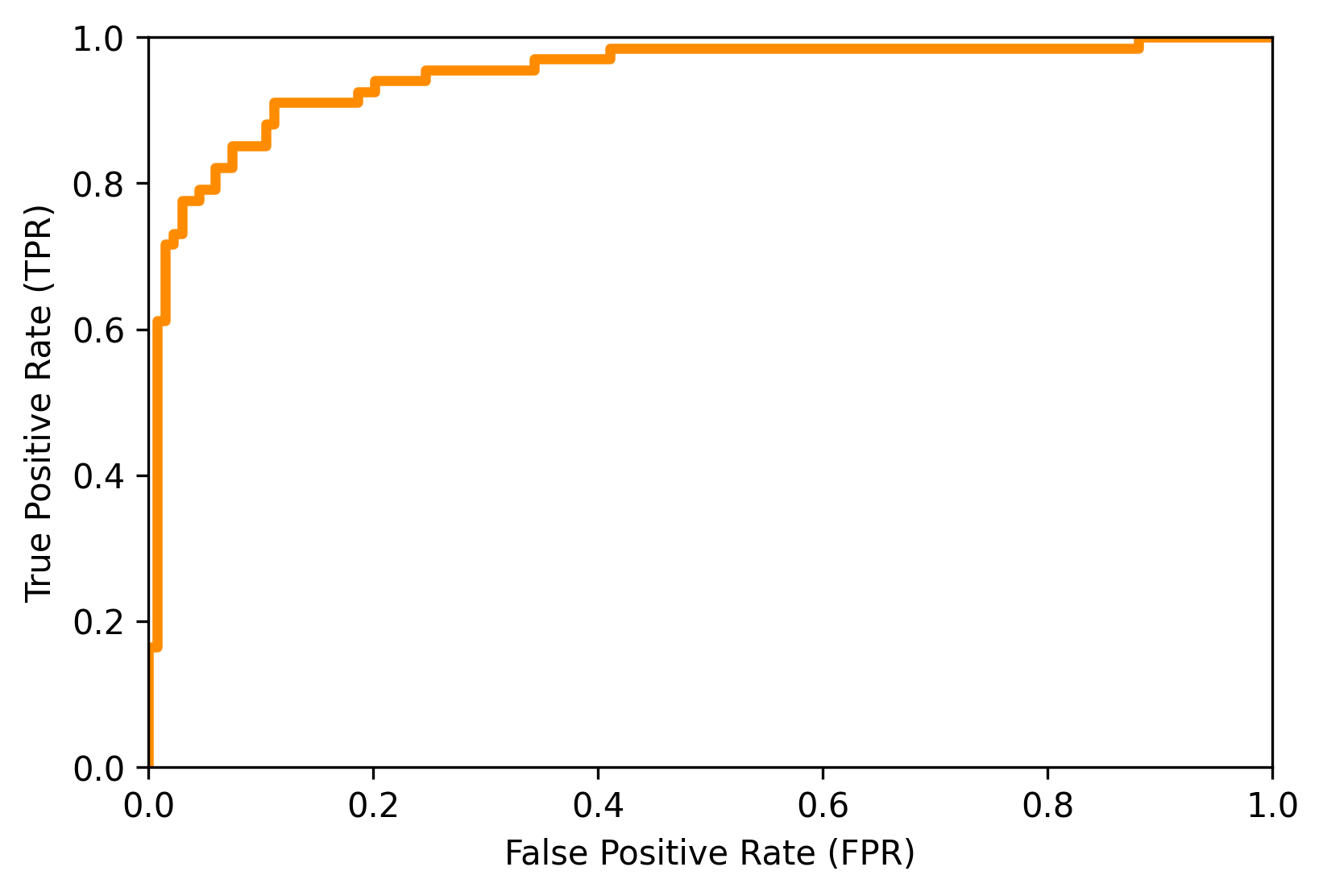}{0.47\textwidth}{\hspace{0.75cm}Neural Network}}
\caption{Receiver Operating Characteristic (ROC) plots for the RF, LR, XGBoost, CatBoost, and NN classifiers on the test set. The x-axis denotes the False Positive Rate (FPR), and the y-axis denotes the True Positive Rate (TPR). The area under the curve AUC metric is the area under the orange curve.}
\label{fig:combinedroc}
\end{figure*}

We also plotted the Receiver Operating Characteristic curve (ROC) to assess the performance of the classifiers. In ROC, we plot the TPR and FPR of a classifier at varying thresholds. A random prediction will generate an Area Under the Curve (AUC) of 0.5, whereas an ideal classifier has an AUC of 1. The ROC plots for all classifiers are shown in Fig. \ref{fig:combinedroc}. All classifiers had very high AUC scores in the range of 0.937-0.961, as listed in Table \ref{tab:modelperf}. AUC metric is considered to be a better measure of performance than comparing accuracy, precision, or recall values as it considers the complete spectrum of classification threshold values.

We have also added the performance metrics of the "Combined Classifier" on the test set in Table \ref{tab:modelperf} for comparison with other classifiers. We observe improved numbers on all the metrics. Again, these results show small uncertainties, i.e., only a minor variation in results was observed across the 100 train/test splits. This implies that there is merit in combining the classifier outputs while still keeping the robustness of results intact. 

\begin{table*}
\caption{Model performance metrics and their uncertainties observed across 100 test sets. Column(1) denotes the type of ML model used. Column(2) lists the overall accuracy observed. Columns(3)-(5) represent the Precision, Recall and F1-score observed for class BL Lac. Columns(6)-(8) represent the Precision, Recall and F1-score observed for class FSRQ. Finally, Column(9) lists the AUC under ROC curve for respective models.}            
\label{tab:modelperf} 
\hskip-1.7cm
\footnotesize
\begin{tabular}{ccccccccc}
\hline\hline
Model & Accuracy & \multicolumn{3}{c}{BLLac} & \multicolumn{3}{c}{FSRQ} & AUC\\
 & & Precision & Recall & F1-score & Precision & Recall & F1-score & \\
\hline
Random  & 0.911 $\pm$ 0.013 & 0.927 $\pm$ 0.014 & 0.941 $\pm$ 0.014 & 0.934 $\pm$ 0.009 & 0.879 $\pm$ 0.025 & 0.852 $\pm$ 0.031 & 0.865 $\pm$ 0.020 & 0.951 $\pm$ 0.016\\
Forest & &&&&&&&\\
Logistic  & 0.894 $\pm$ 0.013 & 0.947 $\pm$ 0.012 & 0.890 $\pm$ 0.019 & 0.918 $\pm$ 0.011 & 0.806 $\pm$ 0.027 & 0.901 $\pm$ 0.024 & 0.850 $\pm$ 0.018 & 0.937 $\pm$ 0.014\\
Regression & &&&&&&&\\
XGBoost  & 0.912 $\pm$ 0.012 & 0.930 $\pm$ 0.014 & 0.938 $\pm$ 0.014 & 0.934 $\pm$ 0.009 & 0.876 $\pm$ 0.024 & 0.859 $\pm$ 0.031 & 0.867 $\pm$ 0.019 & 0.961 $\pm$ 0.016\\
Classifier & &&&&&&&\\
CatBoost  & 0.913 $\pm$ 0.012 & 0.930 $\pm$ 0.013 & 0.940 $\pm$ 0.013 & 0.935 $\pm$ 0.009 & 0.879 $\pm$ 0.023 & 0.860 $\pm$ 0.029 & 0.869 $\pm$ 0.019 & 0.961 $\pm$ 0.015\\
Classifier & &&&&&&&\\
Neural  & 0.904 $\pm$ 0.012 & 0.930 $\pm$ 0.014 & 0.926 $\pm$ 0.015 & 0.928 $\pm$ 0.009 & 0.855 $\pm$ 0.026 & 0.862 $\pm$ 0.032 & 0.858 $\pm$ 0.019 & 0.947 $\pm$ 0.015\\
Network & &&&&&&&\\
\hline
Combined & 0.938 $\pm$ 0.012 & 0.956 $\pm$ 0.011 & 0.952 $\pm$ 0.013 & 0.954 $\pm$ 0.009 & 0.902 $\pm$ 0.024 & 0.910 $\pm$ 0.018 & 0.906 $\pm$ 0.012 & 0.964 $\pm$ 0.014\\
Classifier & &&&&&&&\\
\hline
\end{tabular}
\end{table*}

\begin{figure*}[t!]
\gridline{\fig{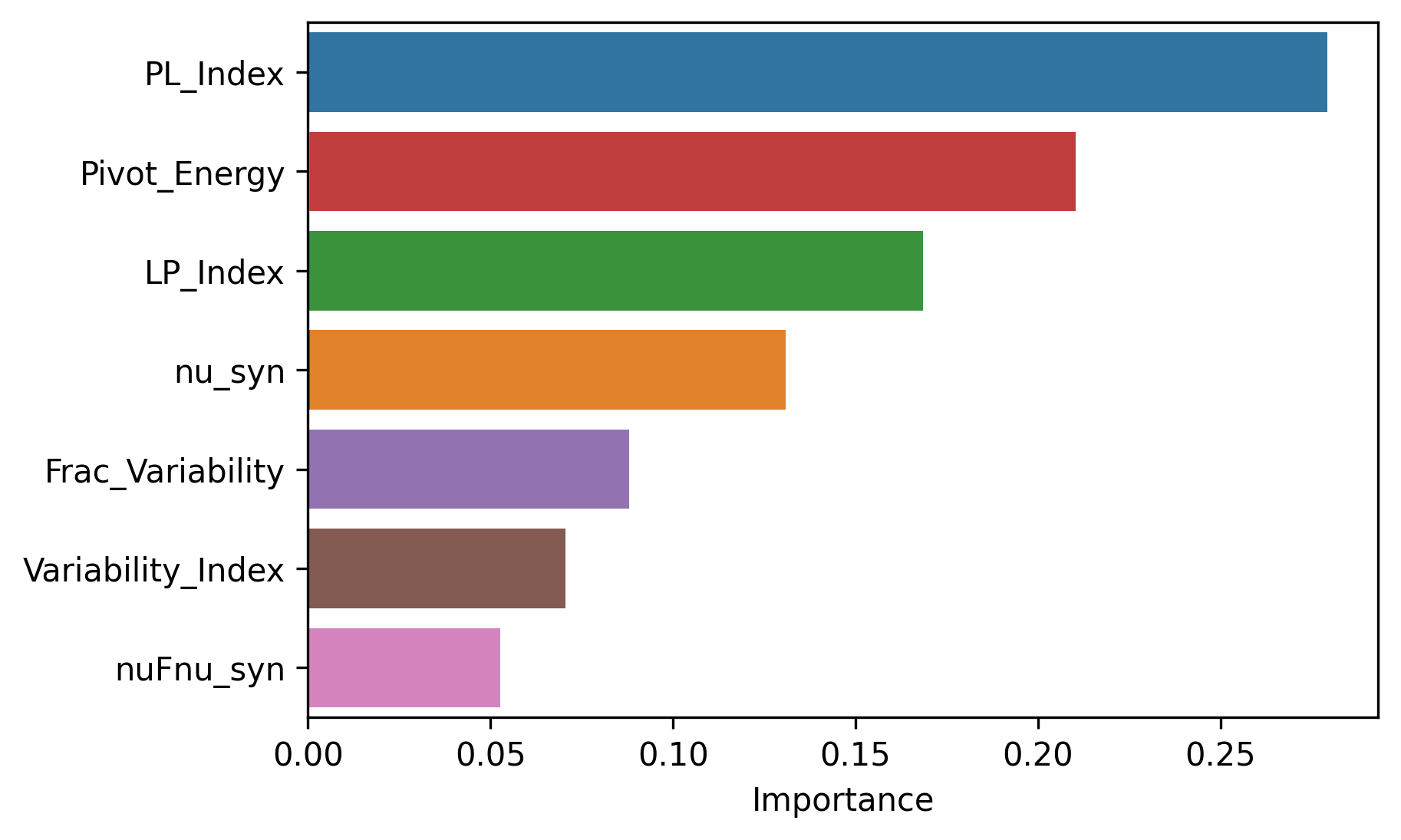}{0.47\textwidth}{\hspace{1.75cm}Random Forest}
          \fig{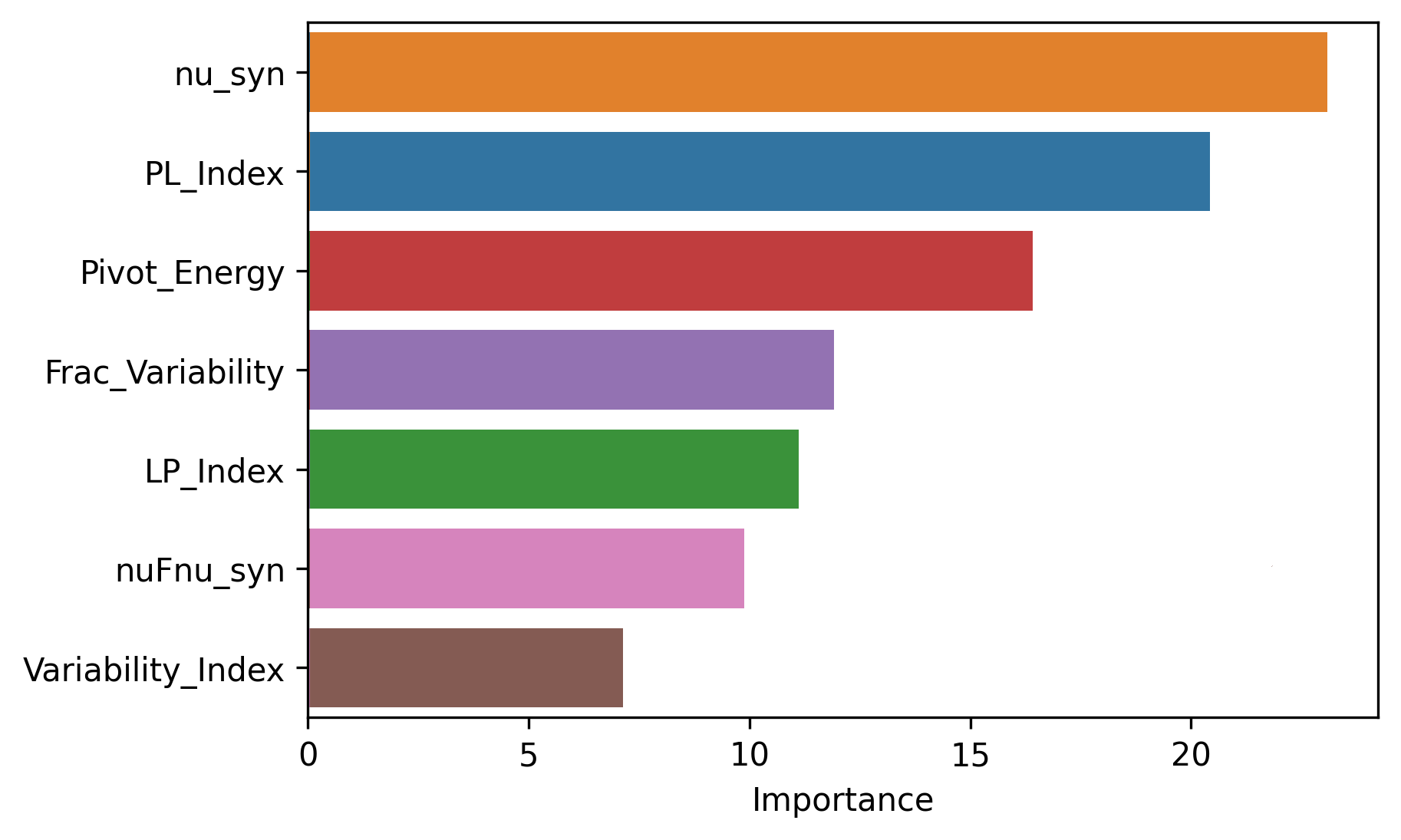}{0.47\textwidth}{\hspace{1.75cm}CatBoost Classifier}}
\gridline{\fig{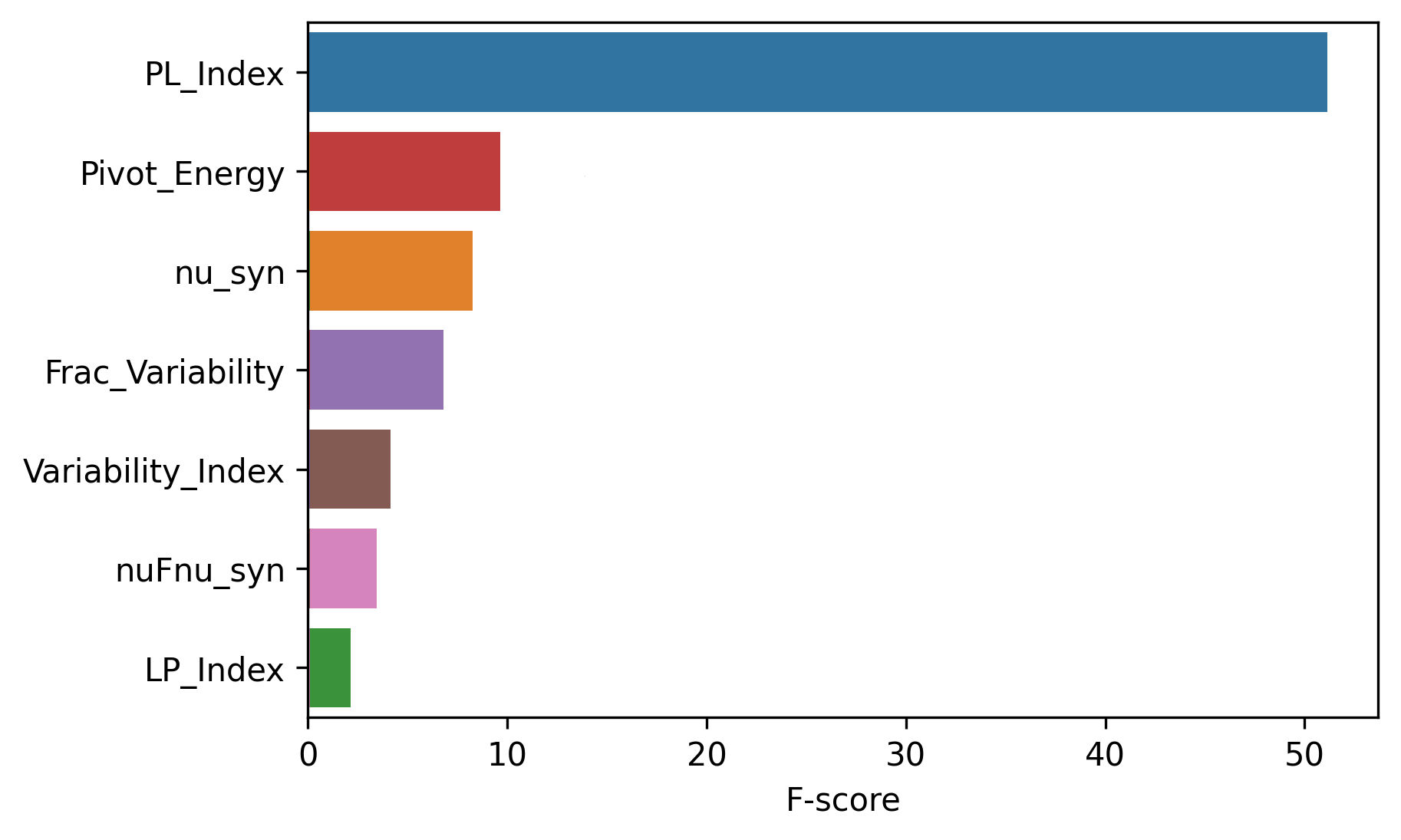}{0.47\textwidth}{\hspace{1.75cm}XGBoost Classifier}}
\caption{Feature Importance as given by the RF, CatBoost, and XGBoost classifiers. The x-axis denotes the feature importance (F-score in the case of XGBoost). The y-axis contains the list of all the features in decreasing order of importance from top to bottom.}
\label{fig:featureimp}
\end{figure*}

We also generated the feature importance from the RF, XGBoost, and CatBoost classifiers for further analysis. The other two classifiers, namely, LR and NN, don't have a systematic way to judge feature importance. The feature importance given by the classifiers is plotted in Fig. \ref{fig:featureimp}. Based on this figure, we find that PL\_index, Pivot\_Energy, and nu\_syn are one of the most important features in the RF plot. These three features are consistent with the feature importance given by the other two methods. Refer to Fig. \ref{fig:allfeaturesimp} to compare the feature importance as given by each classifier as compared to the feature importance calculated while selecting the features.

\begin{figure*}
\centering
\includegraphics[width=1.1\linewidth,clip=true,angle=90,origin=c]{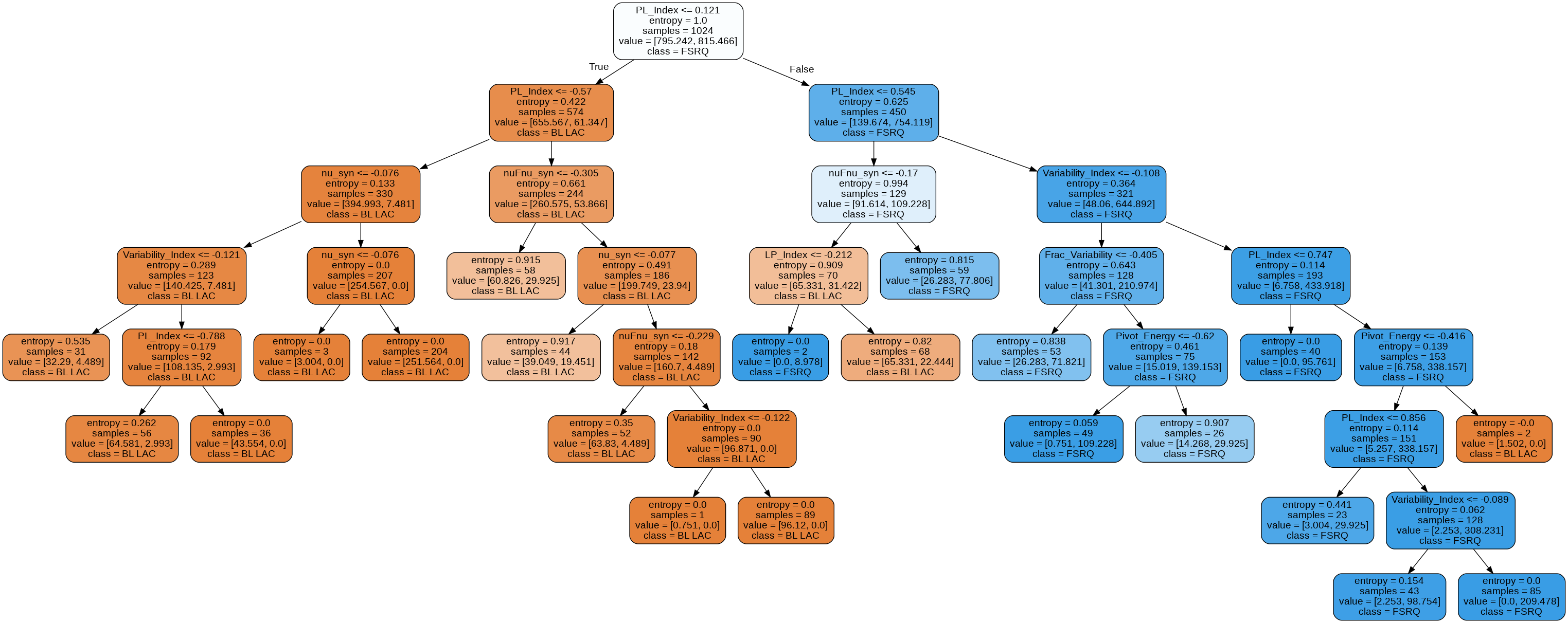}
\caption{\label{fig:treeviz}Tree representation of one of the decision trees trained within RF Classifier. Bluer nodes indicate the node has a larger number of FSRQs, whereas dark orange shades indicate the presence of a larger number of BL Lacs. The first line of all the internal nodes denotes the condition on which the node was split into its children. The left child node contains the sources for which the condition is true, and the right child node contains the sources not satisfying the condition.}
\end{figure*}

In order to make the ML models more interpretable, we diagrammatically represent one of the trees generated as part of the RF Classifier model in Fig. \ref{fig:treeviz}. It shows the tree structure and node composition at various levels in the decision tree. The nodes are color-coded on the scale of orange to blue to show their class composition. Bluer nodes indicate the node has a larger number of FSRQs, whereas darker orange shades indicate the presence of a larger number of BL Lacs. The figure also gives the feature that was used to split the node into its children. For ex., the root node was split based on condition $PL\_index <= 0.121$. The left child node contains the sources for which the condition is true, and the right child node contains the sources not satisfying the condition. Note that all the values shown in the tree are normalized feature values created as part of preprocessing step.

\begin{figure*}[t!]
\gridline{\fig{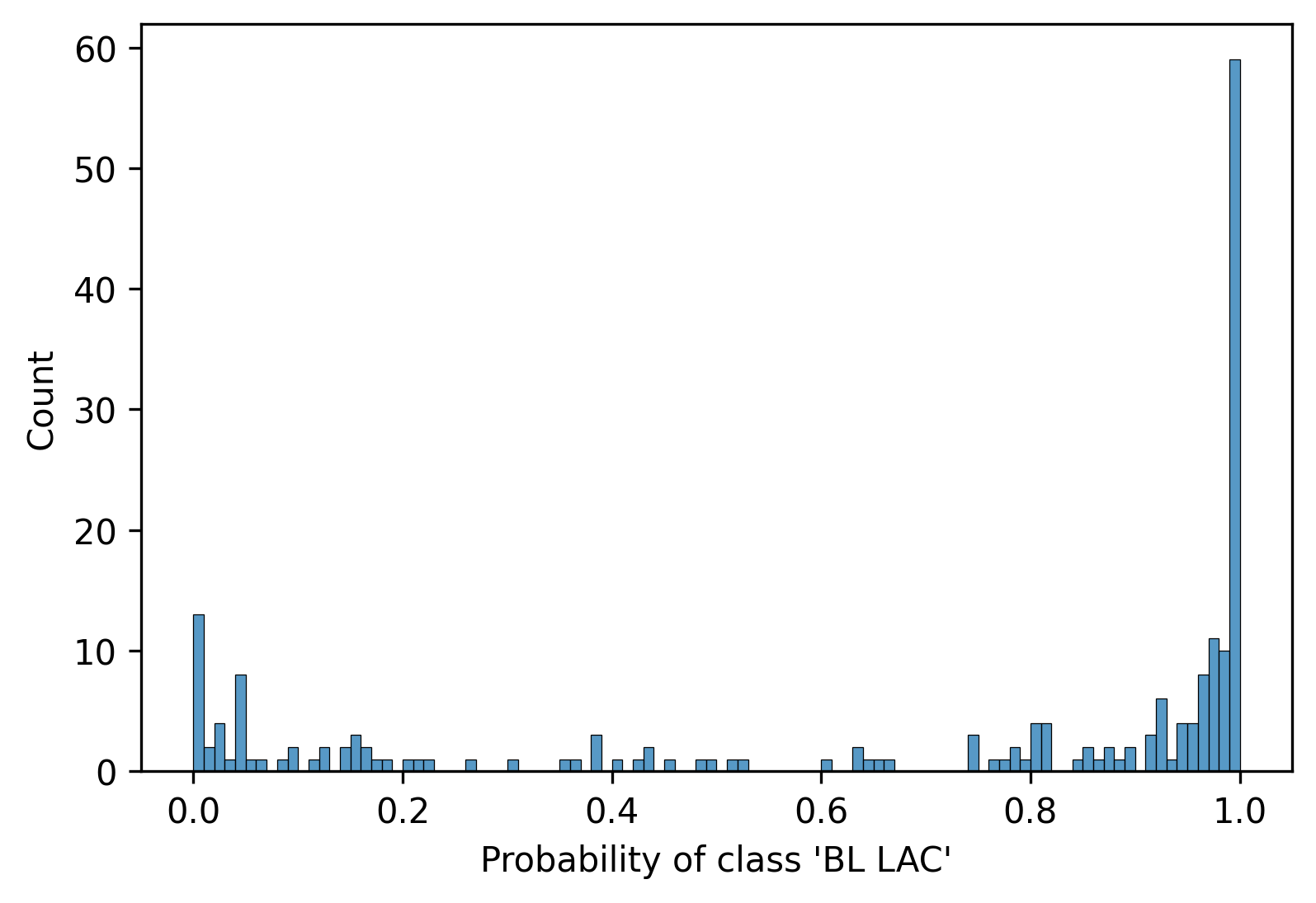}{0.47\textwidth}{\hspace{0.75cm}Random Forest}
          \fig{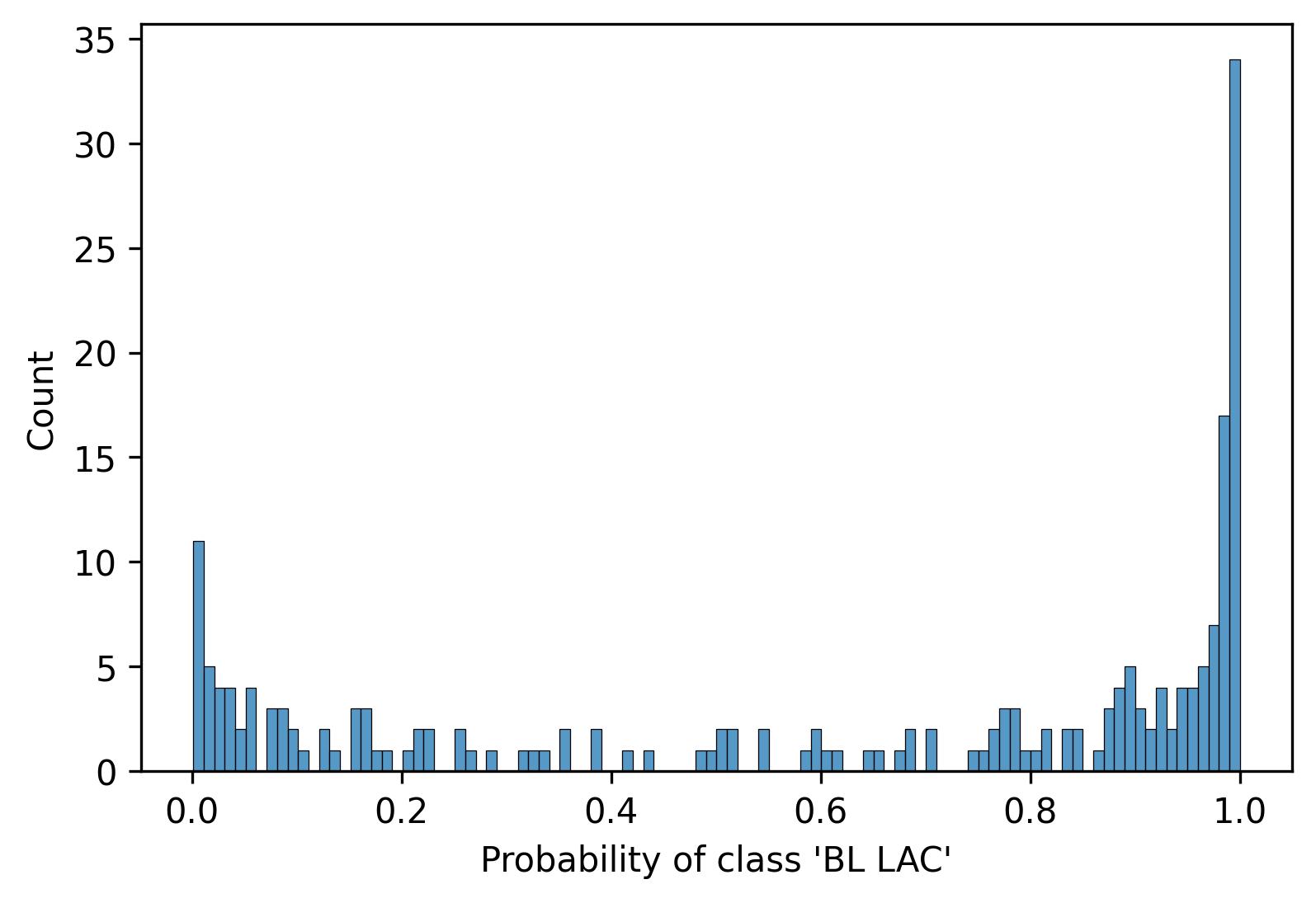}{0.47\textwidth}{\hspace{0.75cm}Logistic Regression}}
\gridline{\fig{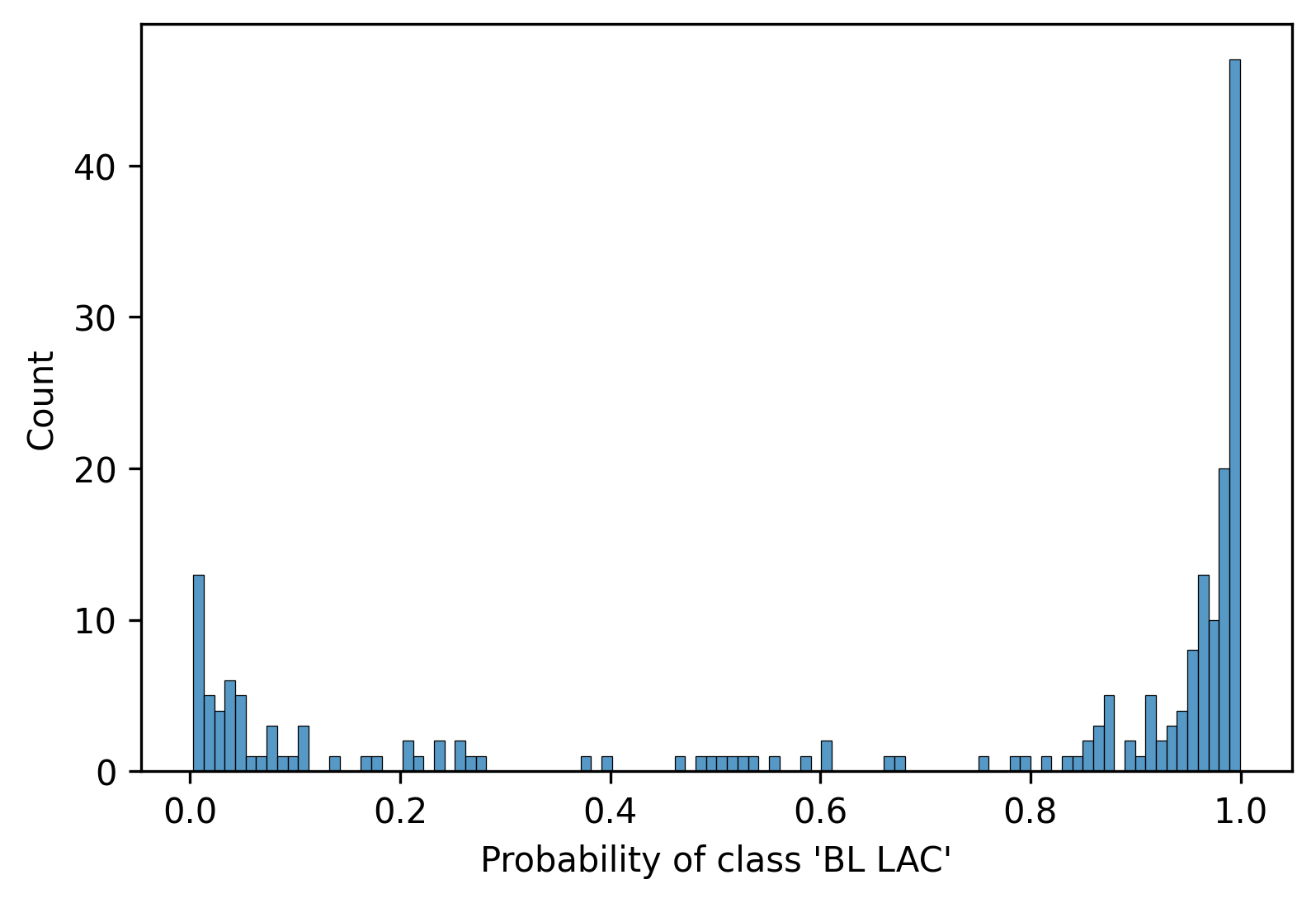}{0.47\textwidth}{\hspace{0.75cm}XGBoost Classifier}
          \fig{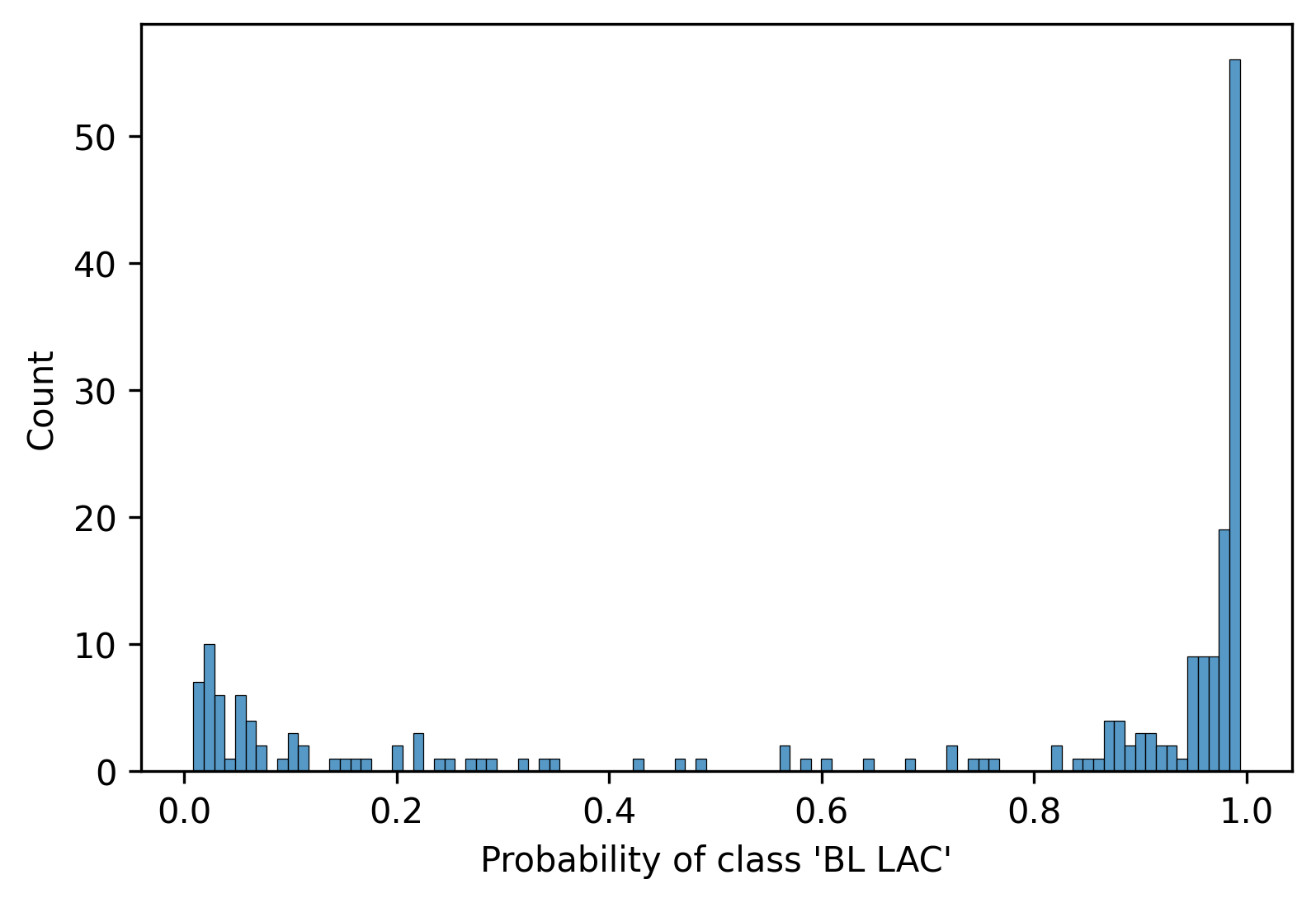}{0.47\textwidth}{\hspace{0.75cm}CatBoost Classifier}}
\gridline{\fig{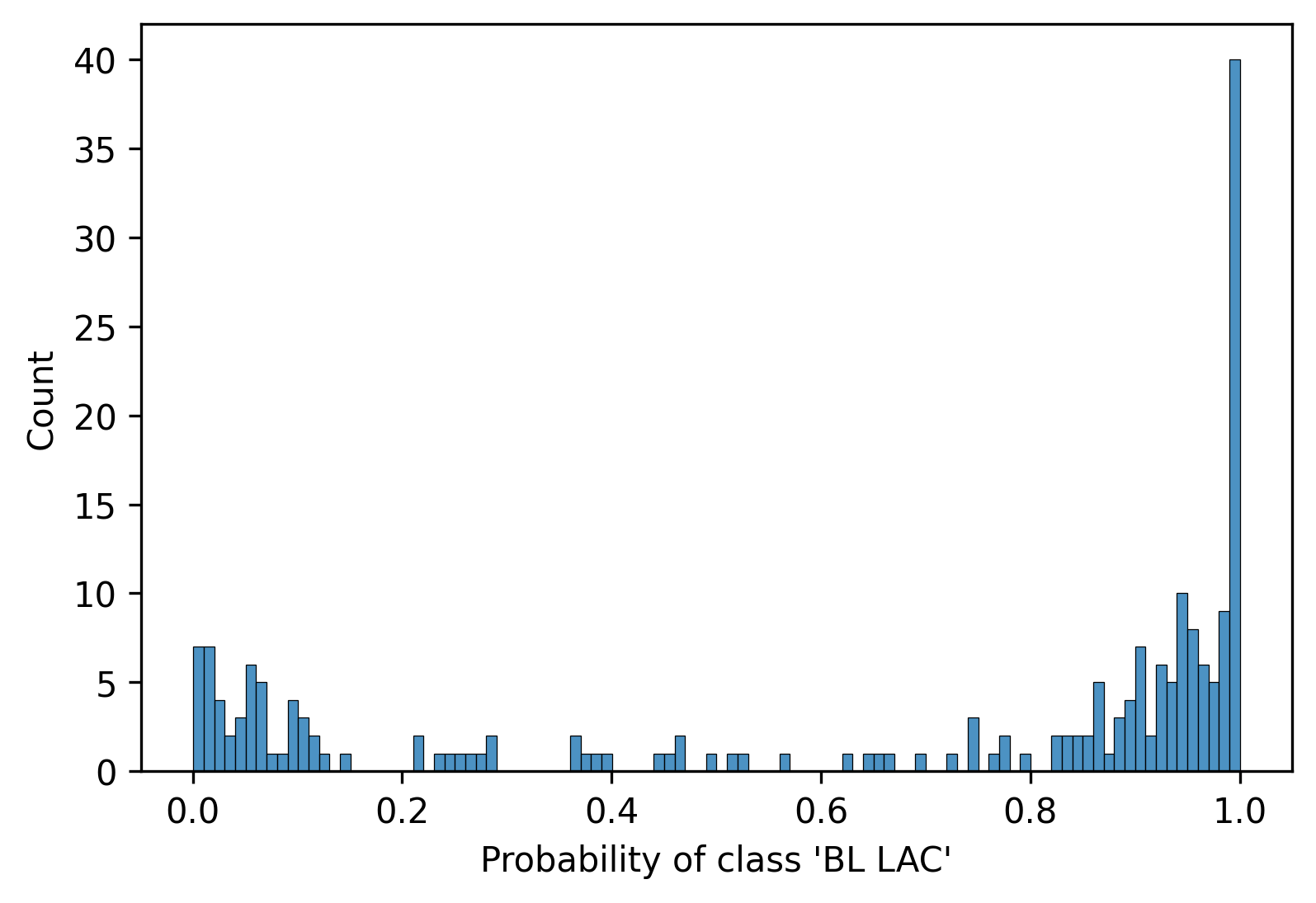}{0.47\textwidth}{\hspace{0.75cm}Neural Network}}
\caption{Probability distribution of the predicted probabilities for different classifiers for class BL Lac on the test set. The x-axis denotes the bins of probabilities in the range [0,1]. The y-axis denotes the count of sources observed in each probability bin.}
\label{fig:testclassprob}
\end{figure*}

Finally, we plot the prediction probabilities for class BL Lac for sources in our test set for all the classifiers. As shown in Fig. \ref{fig:testclassprob}, we observe that most of the sources are concentrated towards the ends of the graph, indicating high class probabilities for either of the two classes. The probabilities $>$ 0.5 indicate the predicted class is BL Lac. Since $P_{bllac} = (1 - P_{fsrq})$, probabilities $<$ 0.5 indicate the predicted class is FSRQ. 

\begin{figure*}[t!]
\gridline{\fig{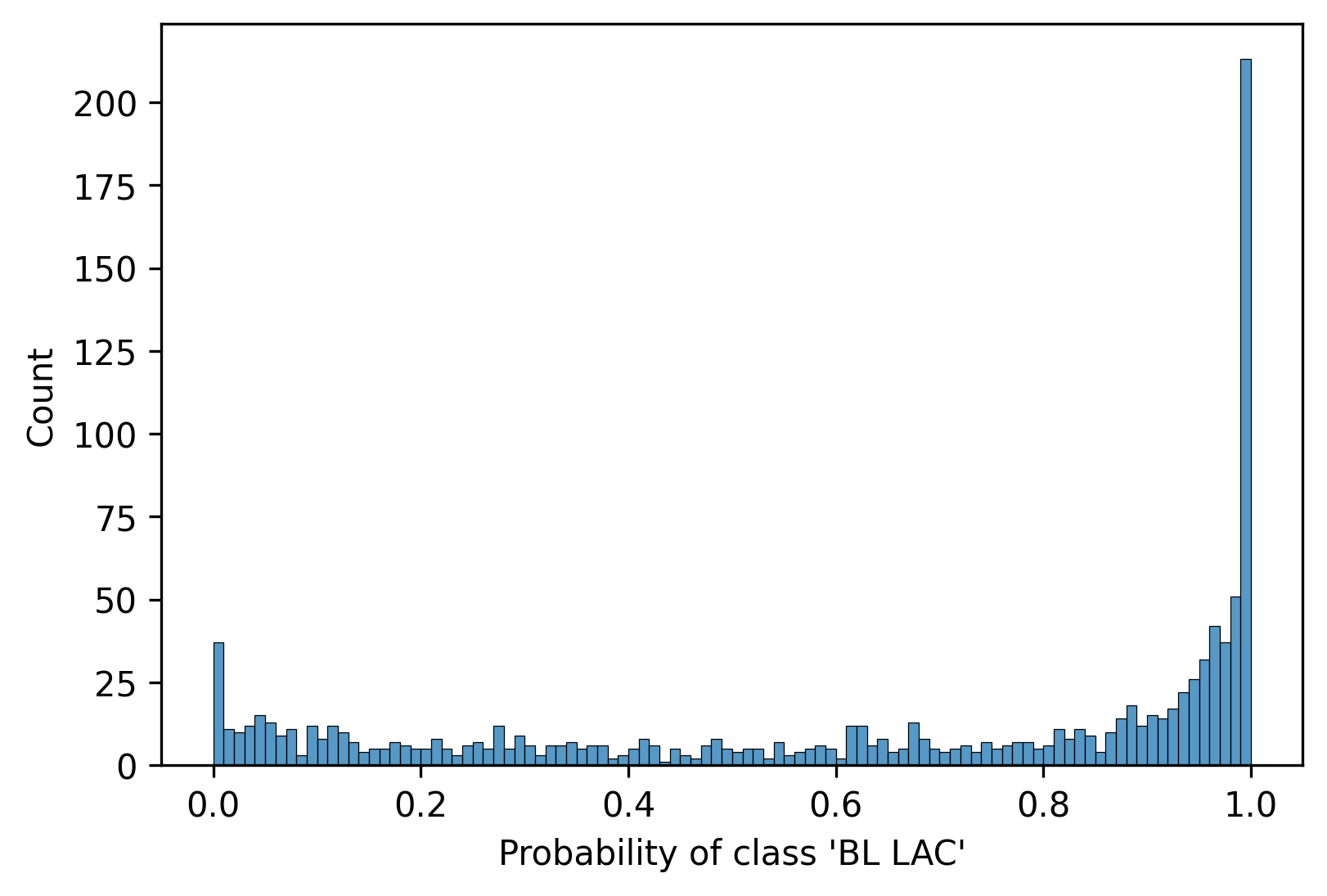}{0.47\textwidth}{\hspace{0.75cm}Random Forest}
          \fig{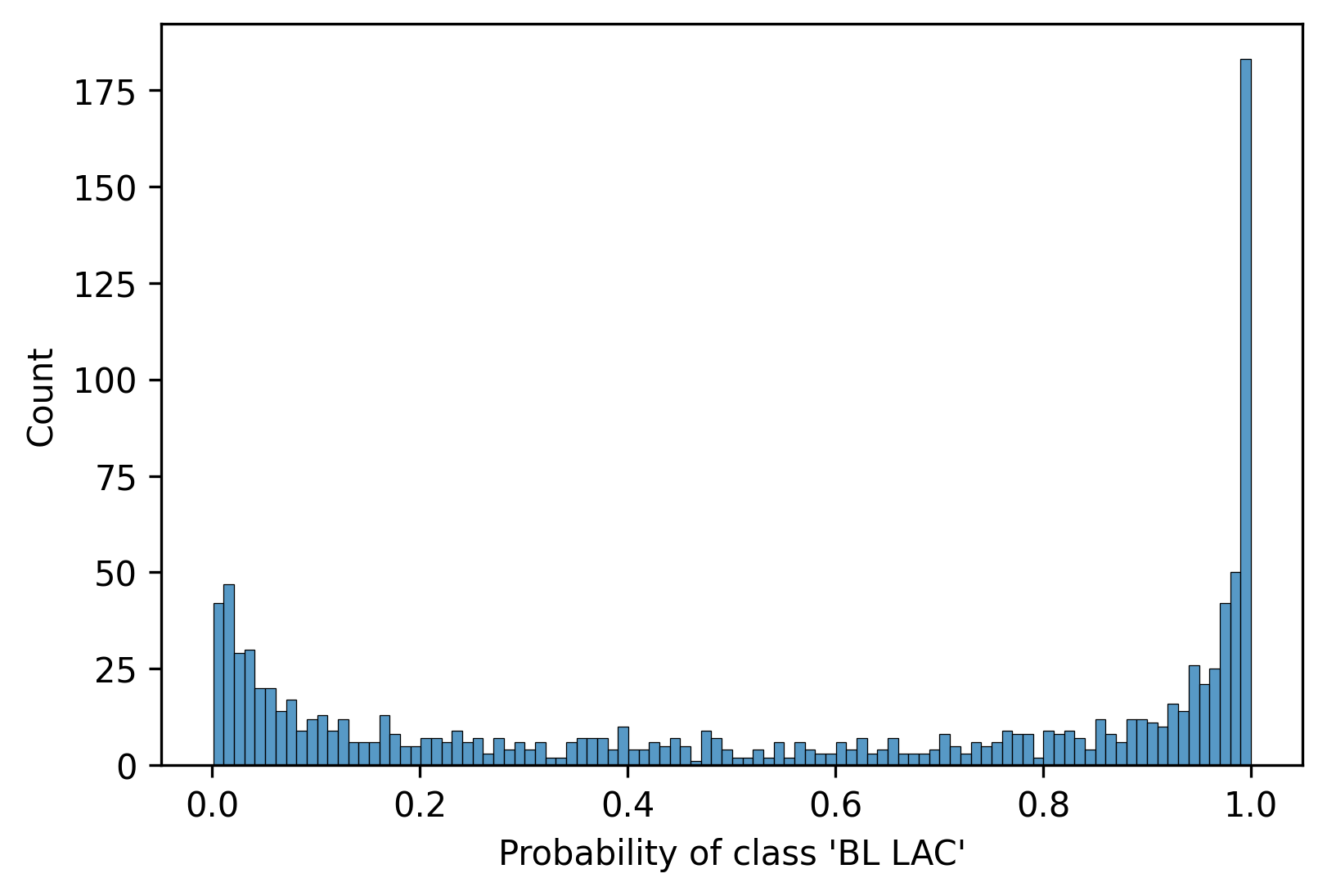}{0.47\textwidth}{\hspace{0.75cm}Logistic Regression}}
\gridline{\fig{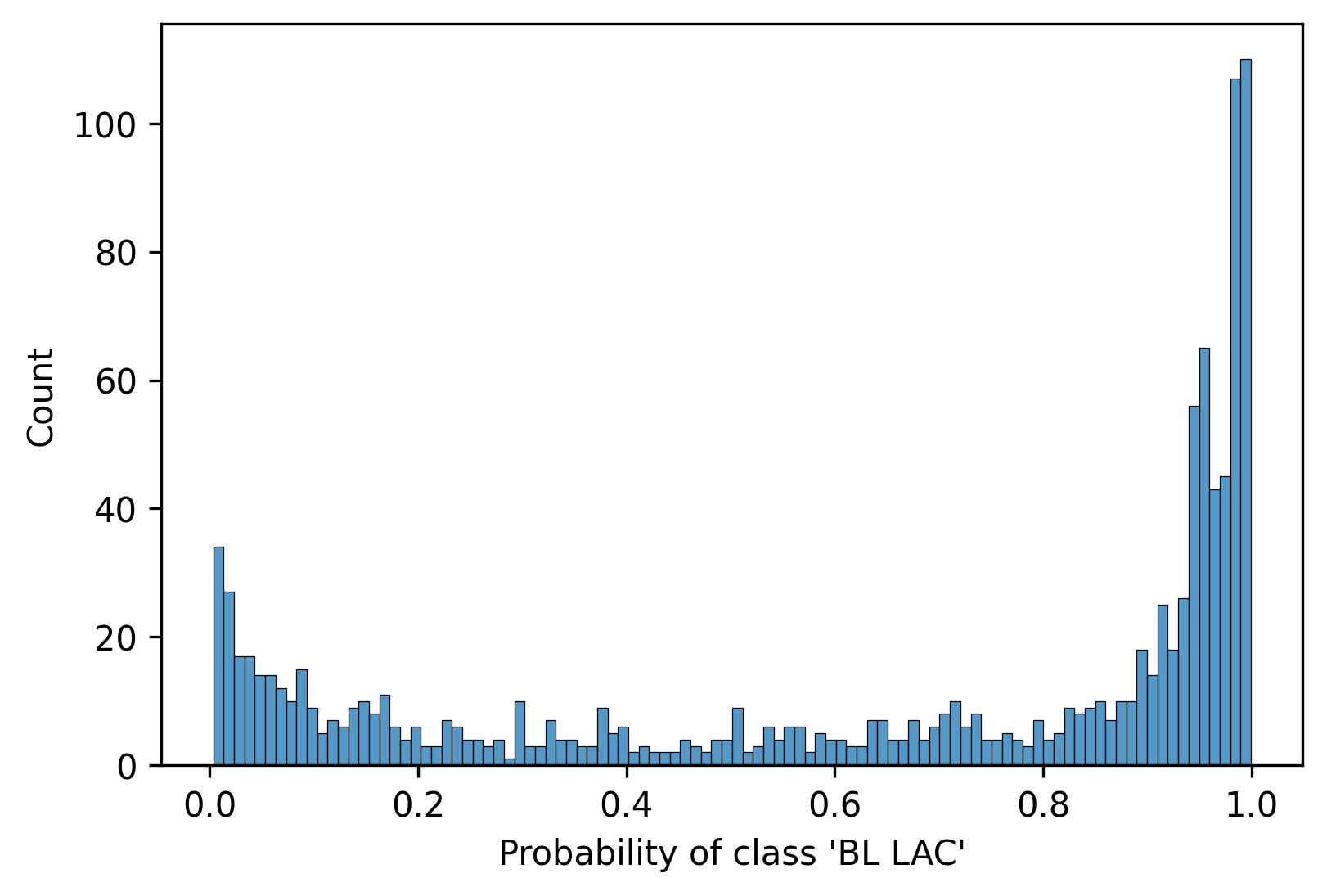}{0.47\textwidth}{\hspace{0.75cm}XGBoost Classifier}
          \fig{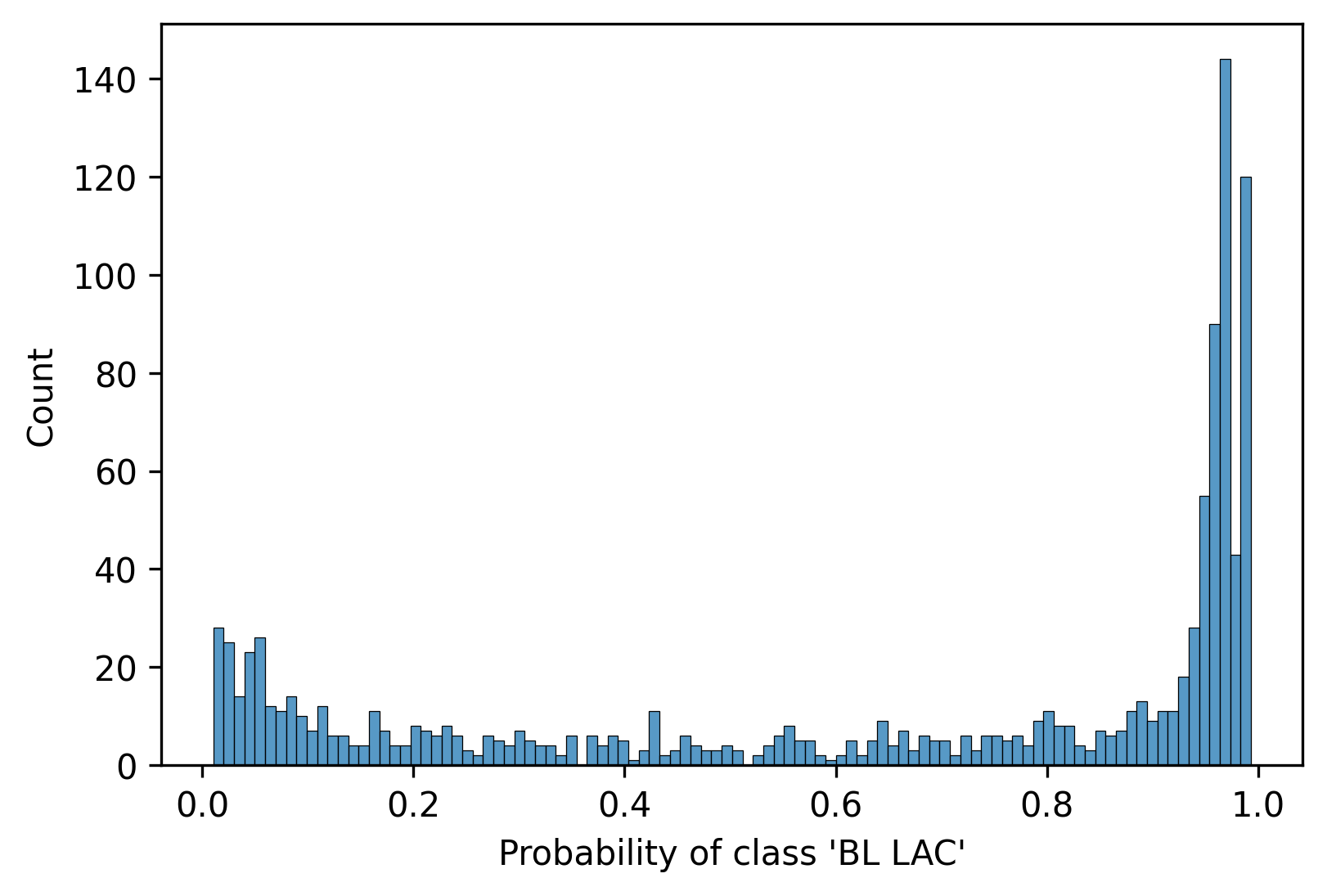}{0.47\textwidth}{\hspace{0.75cm}CatBoost Classifier}}
\gridline{\fig{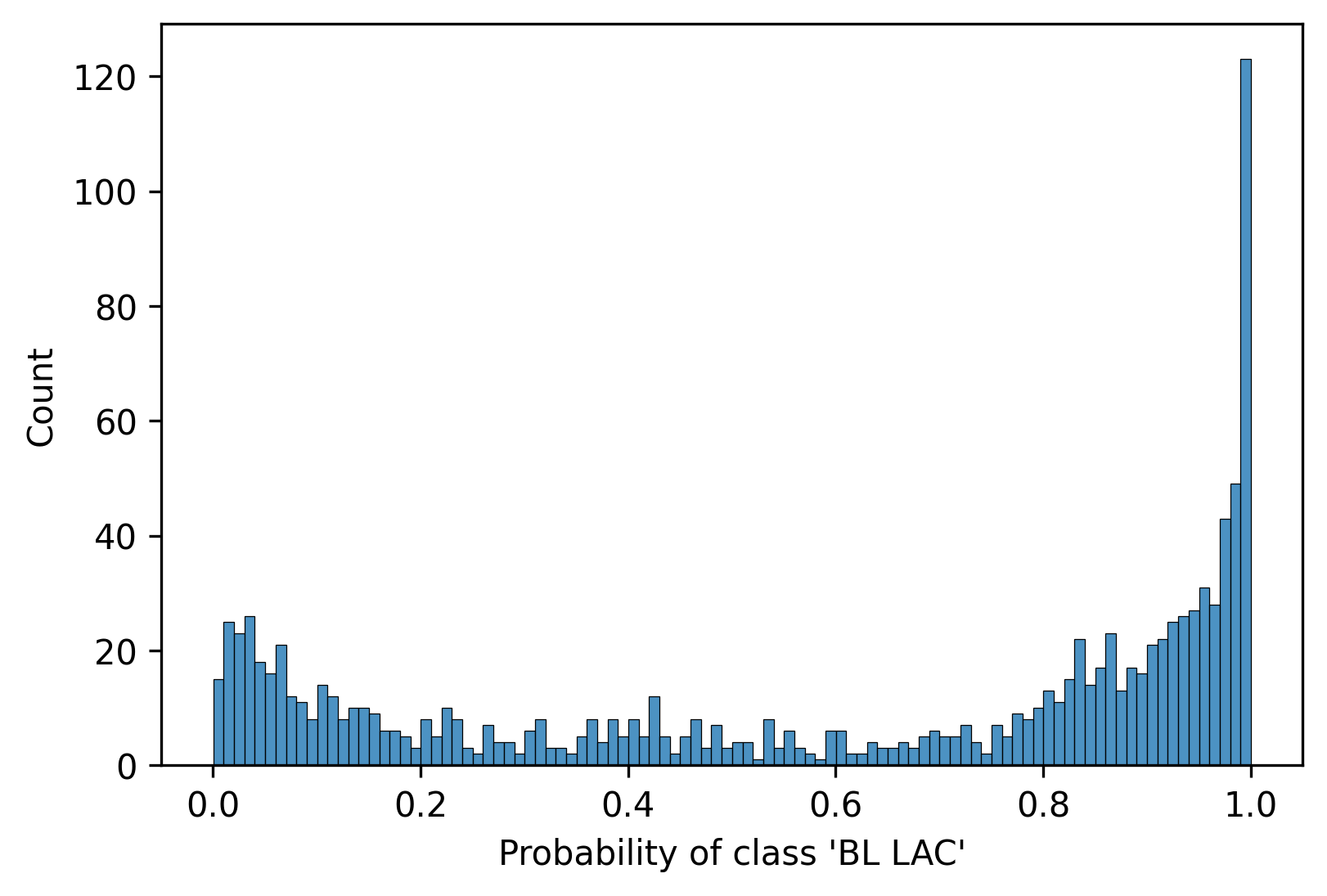}{0.47\textwidth}{\hspace{0.75cm}Neural Network}}
\caption{Probability distribution of the predicted probabilities for different classifiers for class BL Lac on the set of BCU sources. The x-axis denotes the bins of probabilities in the range [0,1]. The y-axis denotes the count of sources observed in each probability bin.}
\label{fig:classprob}
\end{figure*}

These results show the performance of all the classifiers on the test set, i.e., on the sources whose true class is already known. To achieve the goal of this work, we still need to apply our methods to the BCU sources to predict their blazar type. To achieve this, we applied the same preprocessing techniques to the BCU sample as we did on the train and test data sets. We selected the seven features and normalized them by the mean and standard deviation derived from the training set. This ensures that no other data or information leaks into the model outside the training set. We, again, plot the class probabilities for class BL Lac for sources in our BCU data set for all the classifiers. As shown in Fig. \ref{fig:classprob}, we observe that similar to the test set results, most of the sources are concentrated towards the ends of the graph, indicating high class probabilities for either of the two classes.

Each one of the classifiers trained above had excellent performance metrics, but we only classified those sources which have a unanimous prediction across all classifiers. This boosts the quality of our prediction, although the total number of classified sources is lower. All the machine learning algorithms learn different aspects of the data and produce different models even though they are trained on the same data. This is due to the fact that their underlying optimization problem is completely different. For e.g., the same Random Forest algorithm will produce different trees when the splitting criterion is set to 'Gini Impurity' vs. 'Entropy.' This is because the quality of a split calculated is different in both these cases. Hence, a split that is the best for reducing the 'Gini Impurity' may not be the one that reduces the 'Entropy' the most. Similarly, tree-based methods will have a completely different model in comparison to a linear classifier, such as Logistic Regression on the same data split. This brings us to the motivation of combining the results of multiple classifiers and taking a unanimous vote. Each model prediction brings a different perspective on the source.  By combining all the results, we can say that, at a high level, we want to be sure that the classification is consistent from all considered perspectives and that their results agree. To validate it further, we have added the performance metrics of the "Combined Classifier" on the test set in Table \ref{tab:modelperf} for comparison with other classifiers. We observe improved numbers on all metrics. This shows the merit of combining the classifier outputs while keeping the robustness of the results intact.

\begin{figure}
\centering
\includegraphics[width=1.0\linewidth,clip=true]{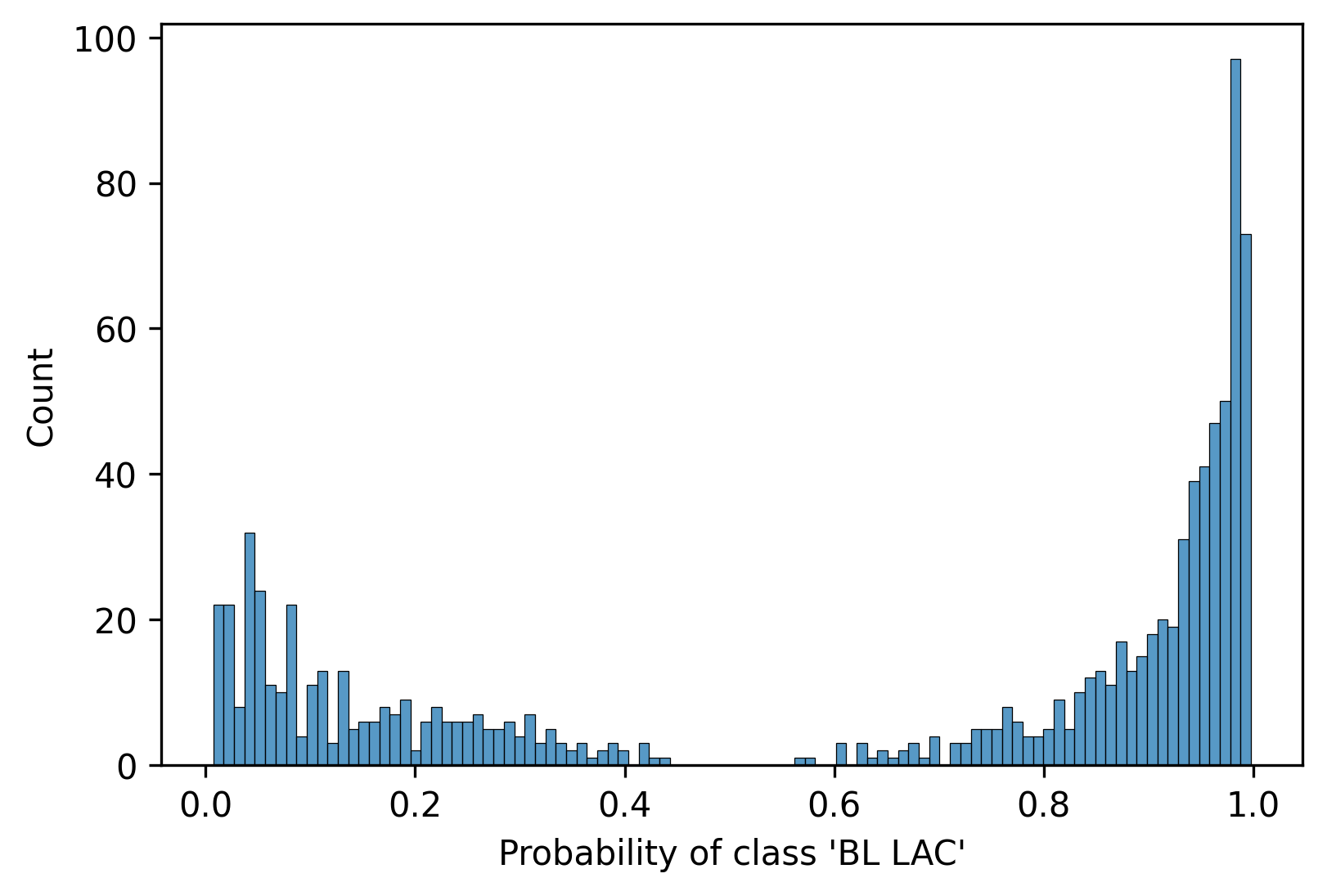}
\caption{Probability distribution of the predicted probabilities for the combined prediction for class BL Lac on the set of BCU sources. The x-axis denotes the bins of probabilities in the range [0,1]. The y-axis denotes the count of sources observed in each probability bin.}
\label{fig:finprobfig}
\end{figure}

 We have plotted the mean probability of our classified BCUs in Fig. \ref{fig:finprobfig}. This clearly shows that most of the classified sources have high average class probabilities as they are concentrated towards the ends of the graph, and the middle portion is practically empty.

After combining the results from all five methods, we classified 943 BCUs from which 610 BCUs are classified BL Lacs and 333 BCUs as FSRQs, while 172 still remain unidentified. The sample of the catalog is given in Table \ref{tab:sample}. A full table is available online in machine-readable form. Although the number of classified candidates may have decreased, the accuracy is improved as combining multiple algorithms gives better classification results \citep{2019ApJ...872..189K}. Most of the BCUs of 4LAC-DR1 are also present in 4LAC-DR3 with better measurements. To determine the number of new BCUs, we cross-matched 4LAC-DR1 with DR3 and found 300 new BCUs. To predict the classification of these new BCUs, we applied all five Supervised Machine Learning (SML) models. Taking the unanimous vote from all five methods, we were able to classify 242 of these sources as 142 BL Lacs and 100 FSRQs leaving only 58 sources unclassified. These new blazars are a step towards a more complete sample of blazar classes. This will help uncover new and extreme physical mechanisms in both classes of blazars.

\section{Discussion and Conclusion}
\label{sect:dis}

Significant progress in periodic updates of 4LAC has led to a larger blazar sample for understanding blazar sequence \citep{1998MNRAS.299..433F} and understanding the physics of high-energy emission from blazar jets. In the present study, we classify the blazar-type candidates of 4LAC-DR3 into two classes of blazars, i.e., BL Lac objects and FSRQs, using supervised ML algorithms, RF, XGBoost, NN, LR, and boost.
Features used for learning the parameters of ML algorithms are Photon index when fitting with PowerLaw (PL\_Index), synchrotron-peak frequency  in the observer frame (nu\_syn), Pivot Energy (Pivot\_Energy, in MeV), Photon index at Pivot\_Energy when fitting with LogParabola (LP\_Index), Fractional variability (Frac\_Variability),  $\nu F\nu$  at synchrotron-peak frequency (nuFnu\_syn, in erg cm$^{-2}$ s$^{-1}$), and Variability index (Variability\_Index). Based on the study of the usefulness of these features on the classification task, i.e., feature importance, it is evident that the Photon index had the most contribution towards it.
It should be highlighted that this study is based on the latest updated data set of the 4LAC catalog and not considering any external data from other data sources or archives. The observational data may cause some bias in the prediction results. However, we noticed that each ML model individually predicted the classification highly accurately. They performed exceedingly well on all the model performance metrics, especially the AUC is reported in the range of 0.937-0.961 for all. Owing to the limitations of the data distribution and ML algorithms, prediction probabilities midway around 0.5 might be unreliable. To further increase the accuracy of our predictions, we label a BCU as BL Lac or FSRQ only when predicted unanimously by all five models. This considerably increased the reliability of the predictions as measured by mean probabilities. 
Combining the prediction results from all five ML algorithms, we present a new catalog of 1115 BCUs classified as 610 BL Lac and 333 FSRQ candidates in Table \ref{tab:sample}. The full table is available online in machine-readable form. According to our study, 172 still remain unclassified, and the determination of their true class needs further investigation. Furthermore, comparing 4LAC-DR1 with DR3 gave us 300 new BCUs. Following the same classification methodology described above, we were able to classify 242 of these candidates as 142 BL Lacs and 100 FSRQs. From these, only 58 BCUs remain unclassified. The larger sample of known blazar candidates provided in this work presents an opportunity to conduct many science cases. One of the most important is the still debated concept of the blazar sequence. Finding multi-frequency data from Radio to X-rays for these still unidentified sources, along with ML algorithms, can unveil more BL Lacs and FSRQs. It will further reduce the incompleteness of the sample and thus offers an interesting perspective for future studies. 

 Owing to powerful telescopes and detectors in the past decade, we can see a dramatic increase in the number of astronomical data sets for millions of sources. Therefore, recently, ML techniques have been widely applied to address the problem of identification and classification of astrophysical sources. Some of the ML algorithms commonly used in Astrophysics include Decision trees, RF, LR, Support Vector Machines (SVM), Artificial Neural Networks (ANN), K-Nearest Neighbours (KNN), Naive Bayes (NB), and Boosted decision trees. In the past, many authors have applied ML algorithms to classify unassociated sources from the $\gamma$-ray Fermi catalogs such as: \citet{2019MNRAS.490.4770K}, \citet{2017MNRAS.470.1291S}, \citet{2019ApJ...887...18K}, \citet{2019ApJ...887..134K}, \citet{2021ApJ...908..177K}, \citet{2021ApJ...923...75K}, \citet{2022arXiv220709307G} and references therein.

Finally, we cross-checked our results with classification predictions of previous works.
First, we compared our results of blazar classification with that of \citet{2019ApJ...887..134K}. They classified Fermi BCUs of the 4FGL catalog using RF, SVM, and ANN, along with optimal combinations of parameters. Based on their analysis, 724 BL Lac type candidates and 332 FSRQ candidates were identified from a sample of 1312 BCUs. The remaining 256 are of the uncertain type for which results were inconsistent. Comparing our total sample set (cleaned) of 1115 BCUs  with the 1312 BCUs of \citet{2019ApJ...887..134K}, we found 812 common BCUs. The non-coincident could be due to improved analysis methods used in 4LAC-DR3, an increase in localization systematic uncertainties, different classifications owing to new spectral properties from the updated catalogs, and recent observations, or due to different algorithms used for predictions, thus updating the results. More details can be found in \citet{2022arXiv220912070T}. From these 812 sources, 600 were examined in both studies and classified with $\sim$ 98.5\% similarity, i.e., 419 were classified as BL Lacs and 172 as FSRQs by both methodologies. In addition to these, we also cross-matched 256 BCUs, which they classified as of uncertain type, and found 154 of them were present in our sample. Of these 154 BCUs, we successfully classified 98 from all five methods as 46 BL Lacs and 52 FSRQs.  Furthermore, from the 303 new BCUs which are not present in their sample set, we are predicting 245 BCUs (143 BL Lacs and 102 FSRQs). We also compared our results with those of \citet{2020MNRAS.493.1926K}, where authors classified 1329 BCUs from the 4FGL catalog into 801 BL Lacs and 406 FSRQs, while 122 remain unclassified. After cross-matching our catalog with them, we found a substantial overlap of 645 BCUs, from which 453 BCUs were classified as BL Lacs and 192 as FSRQs by both studies with $\sim$ 96.1\% similarity. Out of 122 unclassified sources from \citet{2020MNRAS.493.1926K}, we found 75 of them were present in our sample too. Of these 75 BCUs, 46 are now successfully classified as 22 BL Lacs and 24 FSRQs by all five methods, while 29 still remain unclassified. Finally, there are 311 new BCUs that are not present in the sample set of \citet{2020MNRAS.493.1926K}. From these 311 new BCUs, we are now further predicting 251 BCUs as 145 BL Lacs and 106 FSRQs. We also cross-matched our classification results with \citet{universe8080436}. The authors apply SVM to classify BCUs of 4FGL-DR3 to BL Lacs and FSRQs. Comparing our classification with those in \citet{universe8080436}, we found 1114 common BCUs. From these 1114 sources, 547 are classified as BL Lacs and 148 as FSRQs by both studies. Whereas, 5 of our FSRQs are classified as BL Lacs by \citet{universe8080436}, thus giving a $\sim$ 97.98\% similarity. Our results being highly consistent with other ML-based studies provides a piece of evidence that prediction results for the new BCUs are significantly robust. Interested readers can use the parameters given in Table \ref{tab:modelopts} to generate results of this work and can compare the predicted class of BCUs with any other work in the literature. Although each SML model has its own limitations and could lead to some misclassification but applying different algorithms simultaneously proves to be very effective in classifying the unassociated candidates with more precise results.

The primary aim of this work was to continue the classification of unidentified blazar-type sources in the Fermi 4LAC catalog. Classifying each source is a step closer to the goal of identifying the complete gamma-ray sky and having a clear picture of high energy emissions from these sources. It will enable researchers to upgrade the existing theoretical models.
This study will further diversify the blazar sample for a more comprehensive understanding of the blazar and blazar sequence. Moreover, the catalog presented from this work containing BL Lacs and FSRQs will also benefit the community in planning subsequent follow-up spectroscopic observations for not only optical telescopes but also present-day/future multi-frequency observatories such as Cherenkov Telescope Array (CTA), XMM-Newton, Swift, Atacama Large Millimeter/submillimeter Array (ALMA), IceCube, and Imaging X-Ray Polarimetry Explorer. Future observations with better sensitivity will further enable better estimation of redshifts for BL Lacs and FSRQs. For BL Lacs, redshift determination is hampered due to their nearly featureless optical spectra \citep{1995PASP..107..803U}, thus making it an important science case for future blazar studies. To detect weak emission lines in the BL Lacs, we need to obtain high signal-to-noise spectra by utilizing the 8-10m class telescopes \citep{2019ApJ...871..162P}, which is very time-consuming. Because of this, a large fraction of BL Lacs do not have optical counterpart information, and thus most of the unidentified blazar-type candidates in Fermi catalogs are highly likely BL Lacs. The results from the analysis done in this work support the same. There is about a 2:1 ratio between BL Lacs and FSRQs when the output of all five SML algorithms is combined. As pointed out by \citet{2022arXiv220912070T}, photon-index distribution of BCUs in 4LAC-DR3 indicates 2.5 fold increase in the presence of FSRQs in the new BCU sample as compared to 4LAC-DR1 which may be due to a larger flaring tendency of FSRQs as compared to BL Lacs in the Fermi-LAT energy range. A more complete sample of the two mysterious classes of blazars, i.e., BL Lac and FSRQs, is required to gain a better understanding of blazar and its sub-classes.

During the next ten years of time span, the astronomical community is expected to step further into the era of big data when ML algorithms will play a significant role in effectively analyzing and interpreting such voluminous data. Moreover, the upcoming observation facilities can improve the performance of ML algorithms by the addition of new features, thus making ML models more effective tools in the classification of astrophysical sources. These ML-based classifications can then be utilized for numerous science cases for upcoming deep-sky surveys. The work done in this paper based on $gamma$-ray parameters of the two classes of blazars can further be expanded by including multi-wavelength feature sets, such as X-rays, optical, UV, and radio. We plan to address the same in future work.

\section{Acknowledgement}
We are grateful to the anonymous reviewer for their insightful comments and suggestions, which helped us in improving this work. The author would like to thank the Fermi-LAT Collaboration for the
public availability of data.

\startlongtable
\begin{longrotatetable}
\begin{deluxetable*}{ccccccccccccccccccccc}
\centerwidetable
\label{tab:sample}
\tablecaption{An example of classification of Fermi BCUs with ML algorithms. The complete table is available in a machine-readable format.}
\tablewidth{0pt}
\tablehead{
\colhead{Source Name} & \colhead{PLI} & \colhead{PE} & \colhead{LPI} & \colhead{nu\_syn} & \colhead{nuFnu\_syn} & \colhead{VI} & \colhead{FV} &
\colhead{$P_{NN}$} & \colhead{$err_{NN}$} & \colhead{$P_{RF}$} & \colhead{$err_{RF}$} & \colhead{$P_{XG}$} & \colhead{$err_{XG}$} & \colhead{$P_{LR}$} & \colhead{$err_{LR}$} & \colhead{$P_{CB}$} & \colhead{$err_{CB}$} & \colhead{CLASS} & \colhead{$P_{mean}$} & \colhead{$err_{mean}$}}
\tabletypesize{\tiny}
\startdata
J0001.2+4741 & 2.27170 & 2420.575 & 2.25408 & 1.000E+14 & 3.716E-13 & 25.31395 & 0.67588 & 0.759 & 0.045 & 0.924 & 0.035 & 0.934 & 0.021 & 0.416 & 0.027 & 0.936 & 0.015 & ambiguous & 0.794 & 0.007 \\
J0001.8-2153 & 1.87666 & 4429.934 & 1.71662 & 1.660E+13 & 9.716E-13 & 24.55797 & 0.90285 & 0.993 & 0.003 & 0.957 & 0.023 & 0.961 & 0.018 & 0.961 & 0.010 & 0.905 & 0.024 & bllac & 0.955 & 0.004 \\
J0002.3-0815 & 2.09207 & 3399.594 & 2.06073 & 7.586E+13 & 3.993E-13 & 13.01421 & 0.09686 & 0.911 & 0.014 & 0.966 & 0.025 & 0.957 & 0.014 & 0.929 & 0.013 & 0.972 & 0.005 & bllac & 0.947 & 0.003 \\
J0002.4-5156 & 1.91446 & 4073.996 & 1.53527 & 0.000E+00 & 0.000E+00 & 17.68631 & 0.57169 & 0.991 & 0.005 & 0.993 & 0.009 & 0.984 & 0.006 & 0.968 & 0.009 & 0.971 & 0.005 & bllac & 0.981 & 0.002 \\
J0003.1-5248 & 1.91551 & 3392.686 & 1.85940 & 0.000E+00 & 0.000E+00 & 7.99843 & 0.00000 & 0.972 & 0.007 & 0.999 & 0.003 & 0.976 & 0.009 & 0.988 & 0.002 & 0.967 & 0.007 & bllac & 0.980 & 0.001 \\
J0003.3-1928 & 2.28195 & 1021.799 & 2.10295 & 2.291E+13 & 6.415E-13 & 49.87994 & 0.59188 & 0.337 & 0.054 & 0.557 & 0.072 & 0.622 & 0.099 & 0.381 & 0.019 & 0.632 & 0.087 & ambiguous & 0.506 & 0.016 \\
J0003.5+0717 & 2.21727 & 2149.714 & 1.96438 & 3.447E+12 & 1.066E-12 & 10.95177 & 0.00000 & 0.822 & 0.029 & 0.688 & 0.068 & 0.691 & 0.067 & 0.812 & 0.020 & 0.673 & 0.073 & bllac & 0.737 & 0.013 \\
J0007.7+4008 & 2.13957 & 1652.091 & 1.93209 & 0.000E+00 & 0.000E+00 & 35.66249 & 0.51004 & 0.879 & 0.022 & 0.884 & 0.044 & 0.945 & 0.018 & 0.743 & 0.017 & 0.937 & 0.015 & bllac & 0.878 & 0.006 \\
J0008.4+1455 & 2.07887 & 1593.162 & 1.93790 & 0.000E+00 & 0.000E+00 & 51.85556 & 0.65727 & 0.933 & 0.016 & 0.900 & 0.037 & 0.952 & 0.020 & 0.789 & 0.017 & 0.922 & 0.020 & bllac & 0.899 & 0.005 \\
J0009.8+1340 & 2.03457 & 3366.887 & 1.51148 & 1.567E+16 & 5.457E-13 & 11.89411 & 0.00000 & 0.963 & 0.018 & 0.987 & 0.012 & 0.996 & 0.002 & 0.966 & 0.009 & 0.990 & 0.002 & bllac & 0.980 & 0.002 \\
J0010.8-2154 & 2.37858 & 1205.260 & 2.38856 & 0.000E+00 & 0.000E+00 & 11.98323 & 0.06836 & 0.523 & 0.051 & 0.694 & 0.081 & 0.569 & 0.071 & 0.405 & 0.017 & 0.589 & 0.059 & ambiguous & 0.556 & 0.013 \\
J0011.4-4110 & 2.50065 & 1013.242 & 2.48008 & 6.607E+14 & 2.964E-13 & 7.67801 & 0.00000 & 0.405 & 0.054 & 0.711 & 0.070 & 0.796 & 0.095 & 0.200 & 0.014 & 0.821 & 0.059 & ambiguous & 0.587 & 0.014 \\
J0011.8-3142 & 1.94243 & 4122.475 & 1.86731 & 0.000E+00 & 0.000E+00 & 24.58559 & 0.77018 & 0.979 & 0.007 & 0.996 & 0.006 & 0.983 & 0.007 & 0.940 & 0.013 & 0.963 & 0.008 & bllac & 0.972 & 0.002 \\
J0013.4+0950 & 1.95604 & 5697.543 & 2.01828 & 2.065E+16 & 2.094E-12 & 18.24051 & 0.32545 & 0.955 & 0.016 & 0.978 & 0.020 & 0.990 & 0.006 & 0.978 & 0.008 & 0.989 & 0.002 & bllac & 0.978 & 0.003 \\
J0014.3-0500 & 2.37367 & 1642.393 & 2.36200 & 1.096E+13 & 3.973E-13 & 24.51387 & 0.38751 & 0.593 & 0.060 & 0.430 & 0.082 & 0.582 & 0.078 & 0.298 & 0.012 & 0.387 & 0.061 & ambiguous & 0.458 & 0.014 \\
J0014.9+3212 & 2.55602 & 568.577 & 2.45291 & 1.000E+13 & 7.765E-13 & 42.14543 & 0.51269 & 0.048 & 0.014 & 0.034 & 0.020 & 0.020 & 0.007 & 0.051 & 0.004 & 0.042 & 0.010 & fsrq & 0.039 & 0.003 \\
J0017.0-0649 & 2.26517 & 1300.620 & 2.23275 & 6.383E+13 & 1.256E-12 & 40.29398 & 0.39171 & 0.798 & 0.026 & 0.960 & 0.020 & 0.931 & 0.019 & 0.529 & 0.014 & 0.944 & 0.012 & bllac & 0.833 & 0.004 \\
J0019.2-5640 & 2.26854 & 1376.284 & 2.24366 & 2.786E+12 & 3.315E-13 & 109.07163 & 1.07069 & 0.221 & 0.055 & 0.228 & 0.068 & 0.137 & 0.056 & 0.218 & 0.014 & 0.146 & 0.038 & fsrq & 0.190 & 0.011 \\
J0024.4+4647 & 2.76796 & 650.365 & 2.77139 & 0.000E+00 & 0.000E+00 & 31.76876 & 0.60275 & 0.015 & 0.007 & 0.082 & 0.035 & 0.052 & 0.023 & 0.006 & 0.001 & 0.051 & 0.012 & fsrq & 0.041 & 0.004 \\
J0025.7-4801 & 2.26543 & 1465.884 & 1.92869 & 1.905E+13 & 8.276E-13 & 96.93690 & 1.28239 & 0.451 & 0.117 & 0.590 & 0.069 & 0.419 & 0.137 & 0.173 & 0.020 & 0.553 & 0.099 & ambiguous & 0.437 & 0.022 \\
J0029.4+2051 & 1.74450 & 6446.284 & 1.45616 & 4.786E+15 & 4.759E-13 & 11.58067 & 0.00000 & 0.995 & 0.003 & 0.999 & 0.002 & 0.995 & 0.002 & 0.998 & 0.001 & 0.991 & 0.002 & bllac & 0.996 & 0.000 \\
J0032.3-5522 & 2.25028 & 1021.264 & 2.10050 & 0.000E+00 & 0.000E+00 & 308.65665 & 1.49664 & 0.139 & 0.063 & 0.242 & 0.063 & 0.178 & 0.085 & 0.111 & 0.013 & 0.197 & 0.057 & fsrq & 0.174 & 0.014 \\
J0032.5-4724 & 1.69346 & 7187.079 & 1.46502 & 8.318E+15 & 8.322E-13 & 13.09033 & 0.00000 & 0.996 & 0.002 & 1.000 & 0.000 & 0.997 & 0.001 & 0.999 & 0.000 & 0.992 & 0.001 & bllac & 0.997 & 0.000 \\
J0033.9+3858 & 2.29032 & 1130.386 & 2.04489 & 9.333E+12 & 2.154E-13 & 33.33133 & 0.68494 & 0.309 & 0.060 & 0.519 & 0.075 & 0.472 & 0.110 & 0.326 & 0.020 & 0.370 & 0.090 & ambiguous & 0.399 & 0.017 \\
J0034.0-4116 & 2.62482 & 898.741 & 2.62487 & 9.333E+12 & 7.509E-13 & 39.68293 & 0.74233 & 0.032 & 0.012 & 0.102 & 0.041 & 0.073 & 0.031 & 0.019 & 0.002 & 0.054 & 0.015 & fsrq & 0.056 & 0.006 \\
J0035.0-5728 & 2.52042 & 831.286 & 2.38279 & 4.467E+14 & 1.015E-12 & 8.39924 & 0.00000 & 0.362 & 0.056 & 0.676 & 0.081 & 0.687 & 0.118 & 0.175 & 0.014 & 0.746 & 0.070 & ambiguous & 0.529 & 0.017 \\
J0036.8+1431 & 2.28230 & 2112.066 & 2.27944 & 7.079E+12 & 1.425E-13 & 6.90018 & 0.00000 & 0.746 & 0.051 & 0.574 & 0.078 & 0.669 & 0.122 & 0.688 & 0.021 & 0.648 & 0.069 & bllac & 0.665 & 0.017 \\
J0036.9+1832 & 2.43293 & 924.886 & 2.38508 & 0.000E+00 & 0.000E+00 & 97.43484 & 1.12315 & 0.033 & 0.013 & 0.104 & 0.043 & 0.051 & 0.019 & 0.049 & 0.004 & 0.056 & 0.013 & fsrq & 0.058 & 0.005 \\
J0039.1+4330 & 2.05650 & 2315.259 & 1.92409 & 5.623E+14 & 4.943E-13 & 31.70406 & 0.65185 & 0.948 & 0.011 & 0.979 & 0.017 & 0.991 & 0.003 & 0.844 & 0.012 & 0.980 & 0.005 & bllac & 0.948 & 0.002 \\
J0040.9+3203 & 2.35274 & 1124.145 & 2.14544 & 0.000E+00 & 0.000E+00 & 16.09813 & 0.45853 & 0.347 & 0.056 & 0.454 & 0.068 & 0.414 & 0.087 & 0.293 & 0.015 & 0.465 & 0.064 & fsrq & 0.395 & 0.014 \\
J0041.7-1607 & 1.81181 & 5701.249 & 1.79645 & 1.585E+14 & 3.884E-13 & 16.84383 & 0.48939 & 0.990 & 0.004 & 0.996 & 0.007 & 0.993 & 0.004 & 0.991 & 0.003 & 0.988 & 0.002 & bllac & 0.992 & 0.001 \\
J0043.6+2223 & 2.26415 & 2107.967 & 2.24536 & 8.222E+13 & 5.614E-13 & 16.51427 & 0.32660 & 0.837 & 0.022 & 0.952 & 0.021 & 0.930 & 0.018 & 0.588 & 0.019 & 0.955 & 0.010 & bllac & 0.853 & 0.004 \\
J0044.9+4553 & 2.42543 & 1162.396 & 2.36849 & 1.514E+13 & 9.672E-13 & 10.43431 & 0.00000 & 0.460 & 0.048 & 0.515 & 0.095 & 0.568 & 0.091 & 0.346 & 0.016 & 0.475 & 0.073 & ambiguous & 0.473 & 0.016 \\
J0048.6-2427 & 2.10683 & 2285.883 & 2.03474 & 5.623E+14 & 1.178E-12 & 12.73000 & 0.19151 & 0.917 & 0.011 & 0.976 & 0.017 & 0.978 & 0.008 & 0.894 & 0.008 & 0.982 & 0.004 & bllac & 0.949 & 0.002 \\
J0049.5-4150 & 1.97446 & 4066.654 & 1.88675 & 6.457E+15 & 1.640E-12 & 6.00374 & 0.00000 & 0.957 & 0.012 & 0.952 & 0.036 & 0.977 & 0.015 & 0.982 & 0.004 & 0.987 & 0.003 & bllac & 0.971 & 0.004 \\
J0050.8-3428 & 1.91741 & 4711.380 & 1.92181 & 4.266E+14 & 9.322E-13 & 16.09481 & 0.00000 & 0.965 & 0.011 & 0.998 & 0.005 & 0.995 & 0.002 & 0.990 & 0.003 & 0.991 & 0.002 & bllac & 0.988 & 0.001 \\
J0052.9-6644 & 1.97999 & 3471.569 & 2.03556 & 0.000E+00 & 0.000E+00 & 8.99773 & 0.00000 & 0.957 & 0.010 & 0.988 & 0.012 & 0.955 & 0.024 & 0.978 & 0.004 & 0.959 & 0.008 & bllac & 0.967 & 0.003 \\
J0055.1-1219 & 2.32317 & 846.629 & 2.22125 & 2.018E+12 & 4.820E-13 & 170.33379 & 0.83226 & 0.097 & 0.023 & 0.028 & 0.019 & 0.056 & 0.017 & 0.194 & 0.014 & 0.044 & 0.010 & fsrq & 0.084 & 0.004 \\
J0056.6-5317 & 2.17417 & 1848.381 & 1.96291 & 3.350E+14 & 3.900E-13 & 33.36247 & 0.68481 & 0.853 & 0.026 & 0.946 & 0.027 & 0.983 & 0.006 & 0.614 & 0.020 & 0.968 & 0.008 & bllac & 0.873 & 0.004 \\
J0057.0+4101 & 2.11107 & 2476.713 & 1.94652 & 1.799E+14 & 6.126E-13 & 11.09125 & 0.00000 & 0.907 & 0.016 & 0.998 & 0.005 & 0.992 & 0.003 & 0.922 & 0.009 & 0.985 & 0.004 & bllac & 0.961 & 0.002 \\
J0057.3+2216 & 2.15349 & 2115.055 & 2.13527 & 3.631E+12 & 2.356E-13 & 12.00370 & 0.08429 & 0.894 & 0.015 & 0.702 & 0.082 & 0.623 & 0.083 & 0.861 & 0.010 & 0.790 & 0.056 & bllac & 0.774 & 0.013 \\
\enddata
\tablecomments{Column(1): 4FGL Source names. Column(2): PLI~-- Photon index when fitting with PowerLaw (PL\_Index), Column(3): PE~-- Pivot Energy (Pivot\_Energy, in MeV), Column(4): LPI~-- Photon index at Pivot\_Energy when fitting with LogParabola (LP\_Index), Column(5): nu\_syn~-- synchrotron-peak frequency  in the observer frame (nu\_syn), Column(6): nuFnu\_syn~-- $\nu F\nu$ at synchrotron-peak frequency (nuFnu\_syn, in erg cm$^{-2}$ s$^{-1}$), Column(7) VI~-- Variability index (Variability\_Index), Column(8)   FV~-- Fractional variability (Frac\_Variability), Columns(9)-(18): $P_{NN}$, $P_{RF}$, $P_{XG}$, $P_{LR}$, $P_{CB}$~-- Probability of class BL Lac as given by Neural Network(NN), Random Forest(RF), XGBoost(XG), Logistic Regression(LR) and CatBoost(CB) classifiers. $err_{NN}$, $err_{RF}$, $err_{XG}$, $err_{LR}$, $err_{CB}$~-- Std err of prediction. Column(14): CLASS~-- Predicted class of the source \{bllac, fsrq, ambiguous\}. Column(15): $P_{mean}$ and $err_{mean}$: Average predicted class probabilities and their errors across all classifiers. }
\end{deluxetable*}
\end{longrotatetable}

\bibliographystyle{aasjournal}
\bibliography{main}

\end{document}